\documentclass[fleqn,usenatbib]{rasti}

\pdfoutput=1 
\usepackage{newtxtext,newtxmath}
\usepackage[T1]{fontenc}
\DeclareRobustCommand{\VAN}[3]{#2}
\let\VANthebibliography\thebibliography
\def\thebibliography{\DeclareRobustCommand{\VAN}[3]{##3}\VANthebibliography}
\usepackage[utf8]{inputenc}
\usepackage{hyperref}
\usepackage{graphicx}	
\usepackage{amsmath}	
\usepackage{natbib}
\usepackage{enumitem}
\usepackage{url}
\usepackage{multirow}
\usepackage{xspace}
\usepackage{xcolor}
\usepackage{listings}
\usepackage[capitalise,noabbrev]{cleveref}
\usepackage{booktabs}
\usepackage{subcaption}
\usepackage{tikz}
\usetikzlibrary{fit}
\usepackage{placeins}

\usepackage{verbatim}

\definecolor{background}{rgb}{0.95,0.95,0.92}
\definecolor{codeblack}{rgb}{0.1,0.1,0.1}
\definecolor{codegrey}{rgb}{0.3,0.3,0.3}
\definecolor{codered}{rgb}{0.6,0.1,0.1}
\definecolor{codegreen}{rgb}{0.1,0.6,0.1}
\definecolor{codeblue}{rgb}{0.1,0.1,0.6}

\newcommand\YAMLcolonstyle{\color{codered}\mdseries}
\newcommand\YAMLkeystyle{\color{codeblack}\bfseries}
\newcommand\YAMLvaluestyle{\color{codeblue}\mdseries}

\makeatletter

\newcommand\language@yaml{yaml}

\expandafter\expandafter\expandafter\lstdefinelanguage
\expandafter{\language@yaml}
{
  basicstyle=\YAMLkeystyle\ttfamily\footnotesize,
  keywords={true,false,null,y,n},
  keywordstyle=\color{codeblack}\bfseries,
  sensitive=false,
  comment=[l]{\#},
  morecomment=[s]{/*}{*/},
  commentstyle=\color{codegreen},
  stringstyle=\YAMLvaluestyle,
  moredelim=[l][\color{codered}]{\&},
  moredelim=[l][\color{codeblue}]{*},
  moredelim=**[il][\YAMLcolonstyle{:}\YAMLvaluestyle]{:},   
  morestring=[b]',
  morestring=[b]",
  literate =    {---}{{\ProcessThreeDashes}}3
                {>}{{\textcolor{codered}\textgreater}}1     
                {|}{{\textcolor{codered}\textbar}}1 
                {\ -\ }{{\mdseries\ -\ }}3,
  backgroundcolor=\color{background},
  numberstyle=\tiny\ttfamily\color{codegrey},
  numbers=left,
  numbersep=5pt,
  breaklines=true
}

\lst@AddToHook{EveryLine}{\ifx\lst@language\language@yaml\YAMLkeystyle\fi}
\makeatother

\newcommand\ProcessThreeDashes{\llap{\color{cyan}\mdseries-{-}-}}

\newcommand{\python}{\texttt{python}\xspace}
\newcommand{\cosmopower}{\texttt{CosmoPower}\xspace}
\newcommand{\cosmosis}{\texttt{CosmoSIS}\xspace}
\newcommand{\cobaya}{\texttt{Cobaya}\xspace}
\newcommand{\ccl}{\texttt{CCL}\xspace}
\newcommand{\camb}{\texttt{CAMB}\xspace}
\newcommand{\class}{\texttt{class}\xspace}

\newcommand{\montepython}{\texttt{MontePython}\xspace}
\newcommand{\lcdm}{$\Lambda$CDM\xspace}

\iftrue 
\newcommand{\HTJ}[1]{\texttt{\color{orange}[HTJ: #1]}}
\newcommand{\hjdiff}[1]{{\color{orange} #1}}
\newcommand{\ITRH}[1]{\texttt{\color{red}[ITRH: #1]}}

\newcommand{\BB}[1]{\texttt{\color{magenta}[BB: #1]}}

\else
\newcommand{\HTJ}[1]
\newcommand{\hjdiff}[1]
\newcommand{\ITRH}[1]
\newcommand{\BB}[1]
{}

\fi

\title[Framework for cosmology emulation and inference.]{A complete framework for cosmological emulation and inference with \cosmopower}
\author[Jense et al.]{H. T. Jense,$^{1}$\thanks{E-mail: jenseh@cardiff.ac.uk} I. Harrison,$^{1}$ E. Calabrese,$^{1}$ A. Spurio Mancini,$^{2,3}$ B. Bolliet,$^{4,5}$ J. Dunkley,$^{6,7}$ J. C. Hill$^{8}$
\\
$^{1}$School of Physics and Astronomy, Cardiff University, The Parade, Cardiff, Wales CF24 3AA, UK \\
$^{2}$Department of Physics, Royal Holloway, University of London, Egham Hill, Egham, UK \\
$^{3}$Mullard Space Science Laboratory, University College London, Dorking, RH 5 6NT, UK \\
$^{4}$Kavli Institute for Cosmology, University of Cambridge, Madingley Road, Cambridge CB3 0HA \\
$^{5}$DAMTP, Centre for Mathematical Sciences, Wilberforce Road, Cambridge CB3 0WA, UK \\
$^{6}$Joseph Henry Laboratories of Physics, Jadwin Hall,
Princeton University, Princeton, NJ, USA 08544\\
$^{7}$Department of Astrophysical Sciences, Peyton Hall, 
Princeton University, Princeton, NJ USA 08544\\
$^{8}$Department of Physics, Columbia University, New York, NY 10027, USA
}

\date{Accepted XXX. Received YYY; in original form ZZZ}

\pubyear{2024}

\begin{document}

\label{firstpage}
\pagerange{\pageref{firstpage}--\pageref{lastpage}}
\maketitle

\begin{abstract}

\par We present a coherent, re-usable \python framework which further builds on the cosmological emulator code \cosmopower. In the current era of high-precision cosmology, we require high-accuracy calculations of cosmological observables with Einstein-Boltzmann codes. For detailed statistical analyses, such codes often incur high costs in terms of computing power, making parameter space exploration costly, especially for beyond-$\Lambda$CDM analyses. Machine learning-enabled emulators of Einstein-Boltzmann codes have emerged as a solution to this problem and have become a common way to perform fast cosmological analyses. With the goal of enabling generation, sharing and use of emulators for inference, we define standards for robustly describing, packaging and distributing them, and present software for easily performing these tasks in an automated and replicable manner. We provide examples and guidelines for generating your own sufficiently accurate emulators and wrappers for using them in popular cosmological inference codes. We demonstrate our framework by presenting a suite of high-accuracy emulators for the \camb code's calculations of CMB $C_\ell$, $P(k)$, background evolution, and derived parameter quantities. We show that these emulators are accurate enough for both $\Lambda$CDM analysis and a set of single- and two-parameter extension models (including $N_{\rm eff}$, $\sum m_{\nu}$ and $w_0 w_a$ cosmologies) with stage-IV observatories, recovering the original high-accuracy Einstein-Boltzmann spectra to tolerances well within the cosmic variance uncertainties across the full range of parameters considered. We also use our emulators to recover cosmological parameters in a simulated cosmic-variance limited experiment, finding results well within $0.1 \sigma$ of the input cosmology, while requiring typically $\lesssim1/50$ of the evaluation time than for the full Einstein-Boltzmann computation. 
\end{abstract}




\section{Introduction}
In the last two decades, cosmological observations have become a continuous source of ever-tightening constraints on models of the expansion and composition of the Universe. Bounds on cosmological parameters now come from a variety of measurements. Cosmic Microwave Background (CMB) temperature, polarisation and lensing data from satellite and ground-based experiments -- from e.g., \emph{Planck}\footnote{\url{https://www.cosmos.esa.int/web/planck/pla}}~\citep{ planck2016-l06}, the Atacama Cosmology Telescope\footnote{\url{https://act.princeton.edu/}} (ACT,~\citealp{Aiola_2020,Choi_ACT_DR4,Madhavacheril_ACT_lensing,Qu_ACT_lensing}) and the South Pole Telescope\footnote{\url{https://pole.uchicago.edu/public/Home.html}} (SPT,~\citealp{SPT-3G:2022hvq,SPT:2023jql}) -- yield percent-level limits on the parameters of both the standard $\Lambda$CDM cosmological model and some of its possible extensions. Tests of this model will become even more stringent in the next decade with the new, upcoming CMB observatories such as the Simons Observatory\footnote{\url{https://simonsobservatory.org/}} (SO,~\citealp{SO}) and CMB-S4\footnote{\url{https://cmb-s4.org/}}~\citep{CMB-S4:2016ple}.
In addition, statistics of the late-time distribution of matter such as galaxy lensing and clustering add information on cosmological parameters which track the growth of structures caused by the matter and dark energy fields in the local Universe. These come from a number of large-scale-structure surveys -- including the Dark Energy Survey\footnote{\url{https://www.darkenergysurvey.org/}} (DES, ~\citealp{DES:2021wwk}), the Kilo-Degree Survey\footnote{\url{https://kids.strw.leidenuniv.nl/}} (KiDS, ~\citealp{Heymans:2020gsg}), the Hyper Suprime-Cam Survey\footnote{\url{https://hsc.mtk.nao.ac.jp/ssp/survey/}} (HSC, ~\citealp{More:2023knf,Miyatake:2023njf,Sugiyama:2023fzm}) -- which will soon be overtaken by major new experiments such as the Vera C. Rubin Observatory's Legacy Survey of Space and Time\footnote{\url{https://www.lsstdesc.org/}} (LSST, \citealp{LSSTDarkEnergyScience:2018jkl}), the \emph{Euclid} satellite\footnote{\url{https://www.euclid-ec.org/}} \citep{Euclid:2021icp}, the Nancy Grace Roman Space Telescope\footnote{\url{https://roman.gsfc.nasa.gov/}} \citep{Eifler:2020vvg} and the SPHEREx Observatory\footnote{\url{https://www.jpl.nasa.gov/missions/spherex}} \citep{SPHEREx:2014bgr}. Finally, the imprint of cosmic perturbations on the baryonic matter is mapped by spectroscopic galaxy surveys -- from the Baryon Oscillation Spectroscopic Survey (BOSS,~\citealp{eBOSS:2020yzd}) to the new Dark Energy Spectroscopic Instrument (DESI,~\citealp{DESI:2016fyo, DESI:2024mwx}).
 
The high precision available from these experiments sets strong demands on the accuracy of theoretical modelling of their data vectors, in particular for the upcoming next-generation of surveys, usually labelled as Stage-IV experiments. Higher levels of physical and numerical accuracy in the codes which predict observables in a given cosmological model typically come at the expense of longer evaluation times. Full cosmological exploitation of the data relies on many such evaluations of this `forward model' when calculating likelihood values (and subsequent posterior estimates) and the total amount of time required can easily become intractable. 

\par Various works (see e.g.,~\citealp{CosmoPower2021} and references therein) presented methods to speed up this process by means of emulating the Einstein-Boltzmann codes (typically \camb\footnote{\url{https://github.com/cmbant/CAMB}},~\citealp{Lewis:1999bs}, or \class\footnote{\url{https://github.com/lesgourg/class_public}},~\citealp{lesgourgues2011cosmic,Blas_2011}, which are commonly used to accurately compute linear-theory cosmological power spectra and background evolution quantities). 

However, at present, each study typically generates its own emulator, tailored and limited to the specific need of the analysis performed. This means that there is limited general use for the wide variety of emulators currently existing, due to the lack of standardisation and cross-platform support for them. Aside from applicability and redundancy issues, the ad-hoc use emulators could also be a potential cause of inconsistencies between different analyses, and a  limitation for model comparisons and for data combinations. There are many compelling reasons to expect that further advances in our understanding of cosmology will necessarily come from cross-correlation analyses between different experiments. Such analyses maximise both statistical constraining power and robustness to instrument and astrophysical systematics. They also require the possibility to analyse the different data using a single unified theoretical framework, in order to make consistent predictions for the different data types during inference. Though packages such as \cobaya\footnote{\url{https://github.com/CobayaSampler/cobaya}}\citep{Cobaya2019,Cobaya2021}, \cosmosis\footnote{\url{https://github.com/joezuntz/cosmosis}}~\citep{Zuntz:2014csq} and MontePython~\citep{Brinckmann:2018cvx,Audren:2012wb} exist to enable this, there are still gaps which prevent data combinations which would otherwise be fruitful. By making emulators portable across platforms and frameworks, the work presented here will enable novel data combinations much more easily.

Lack of consistency across emulators also limits the ability to deploy these techniques for other applications. For example, a further use for emulators of Einstein-Boltzmann codes lies in the possibility of \emph{autodifferentiation} (\emph{autodiff.}), a computational method of quickly evaluating partial derivatives of the outputs with respect to their inputs. When these derivatives are known, more effective sampling methods such as Hamiltonian Monte Carlo (HMC), which requires accurately knowing the derivatives of often complex relations, become trivial to include. As an example,~\cite{JaxCosmo} presented a computational framework to autodifferentiate forward models for various cosmological observables. In their paper, they showed how using a specific implementation of HMC known as a No U-Turn Sampler (NUTS) can lead to statistical constraints similar to classical MCMC algorithms in $1/5$th of the time. While for classical Einstein-Boltzmann codes, finding these derivatives is a complicated if not impossible task, this becomes a trivial option when using emulators such as neural networks, in which the computational models used to map between inputs and outputs consist of multiple trivially differentiable units. This means commonly-used software libraries from the recent machine learning revolution, such as \texttt{tensorflow} or \texttt{jax}, are intrinsically able to take advantage of autodiff. \cite{piras2023cosmopowerjax} also recently presented an example of the advantages of combining autodifferentiable emulators with HMC posterior sampling, achieving speed ups $O(10^3)$ relative to traditional Boltzmann codes combined with nested sampling methods.

\begin{figure*}
    \begin{tikzpicture}[node distance=1cm]
        \tikzstyle{block} = [rectangle, rounded corners, minimum width=2.5cm, minimum height=0.6cm, text centered, draw=black]
        
        \node (packaging) [block, fill=red!30, label=above:{\hyperref[subsec:emulated-quantities]{\S\ref{subsec:emulated-quantities}}},minimum width=2.9cm] {Package Prescription};
        \node (generate) [block, fill=green!30, right of=packaging, xshift=2.2cm, label=above:{\hyperref[subsec:emulator-specs]{\S\ref{subsec:training-data}}, \hyperref[subsec:generate-training]{\ref{subsec:generate-training}}}] {Generate Training Data};
        \node (boltzmann-code) [block, fill=blue!30, below of=generate] {Boltzmann code};
        \node (training) [block, fill=green!30, right of=generate, xshift=2cm, label=above:{\hyperref[subsec:network-training]{\S\ref{subsec:network-training}}, \hyperref[subsec:networks-specs-training]{\ref{subsec:network-specs-training}}}] {Train Emulators};
        \node (emulators) [block, fill=red!30, right of=training, xshift=2cm] {Emulator Files};
        \node (validation) [block, fill=green!30, right of=emulators, xshift=2cm, label=above:{\hyperref[subsec:accuracy-validation]{\S\ref{subsec:accuracy-validation}}, \hyperref[subsec:accuracy-plotting]{\ref{subsec:accuracy-plotting}}}] {Accuracy Validation};

        \node (tmp1) [inner sep=-1pt,outer sep=0pt, below of=emulators] {};
        \node (tmp2) [inner sep=-1pt,outer sep=0pt, below of=boltzmann-code] {};
        \node (tmp3) [inner sep=-1pt,outer sep=0pt, below of=tmp1] {};

        \node (wrapper) [block, fill=red!30, below of=tmp2, label=below:{\hyperref[sec:wrapper-description]{\S\ref{sec:wrapper-description}}}] {Wrapper};
        \node (mcmc-software) [block, fill=red!30, below of=wrapper, yshift=-0.5cm] {Sampling Software};
        \node (mcmc-config) [block, fill=red!30, left of=mcmc-software, xshift=-2.2cm,minimum width=2.9cm] {Sampling Configuration};
        \node (mcmc-chain) [block, fill=green!30, right of=mcmc-software, xshift=2cm] {MCMC Chain};
        \node (cosmo-data) [block, fill=blue!30, below of=mcmc-chain] {Cosmological data};
        \node (mcmc-results) [block, fill=green!30, right of=mcmc-chain, xshift=2cm, label=below:{\hyperref[sec:recover-cosmology]{\S\ref{sec:recover-cosmology}}}] {Cosmological constraints};

        \draw [thick,->,>=stealth] (packaging) -- (generate);
        \draw [thick,->,>=stealth] (generate) -- (training);
        \draw [thick,->,>=stealth] (boltzmann-code) -- (generate);
        \draw [thick,->,>=stealth] (training) -- (emulators);
        \draw [thick,->,>=stealth] (emulators) -- (validation);

        \draw [thick,->,>=stealth,draw=red] (packaging) -- (mcmc-config);

        \draw [thick,->,>=stealth] (mcmc-config) -- (mcmc-software);
        \draw [thick,->,>=stealth] (mcmc-software) -- (mcmc-chain);
        \draw [thick,->,>=stealth] (cosmo-data) -- (mcmc-chain);
        \draw [thick,->,>=stealth] (mcmc-chain) -- (mcmc-results);

        \draw [thick,->,>=stealth,draw=red] (emulators) -- (tmp3) -- (tmp2) -- (wrapper);
        \draw [thick,->,>=stealth,draw=black] ([shift={(0.5,0)}]wrapper.south) -- ([shift={(0.5,0)}]mcmc-software.north);
        \draw [thick,->,>=stealth,draw=black] ([shift={(-0.5,0)}]mcmc-software.north) -- ([shift={(-0.5,0)}]wrapper.south);

        \node[draw=blue,inner sep=5mm,fit=(packaging) (boltzmann-code) (emulators) (validation),label=above:\textit{Emulator creation}] {};
        \node[draw=blue,inner sep=5mm,fit=(mcmc-config) (mcmc-results) (cosmo-data) (wrapper),label=below:\textit{Emulator usage}] {};
    \end{tikzpicture}
    \caption{An overview of the workflow with \cosmopower: To create a new emulator (\textbf{top blue box}), we write a \emph{packaging prescription}, use that to \emph{generate training data}, and from that \emph{train emulators} which outputs several \emph{emulator files}, for which we can easily generate plots which \emph{validate the accuracy} of the emulators. This packaging prescription and set of emulator files are then shared with the end-user (\textbf{red arrows}), who wants to use the emulators (\textbf{bottom blue box}): the prescription is put inside the \emph{sampling configuration file}, which is given to our \emph{software wrappers}, which provide the user with an \emph{MCMC chain} that can be used to find \emph{cosmological constraints}. The various labels refer to the sections where these individual steps are described in this paper.} \label{fig:scheme}
\end{figure*}
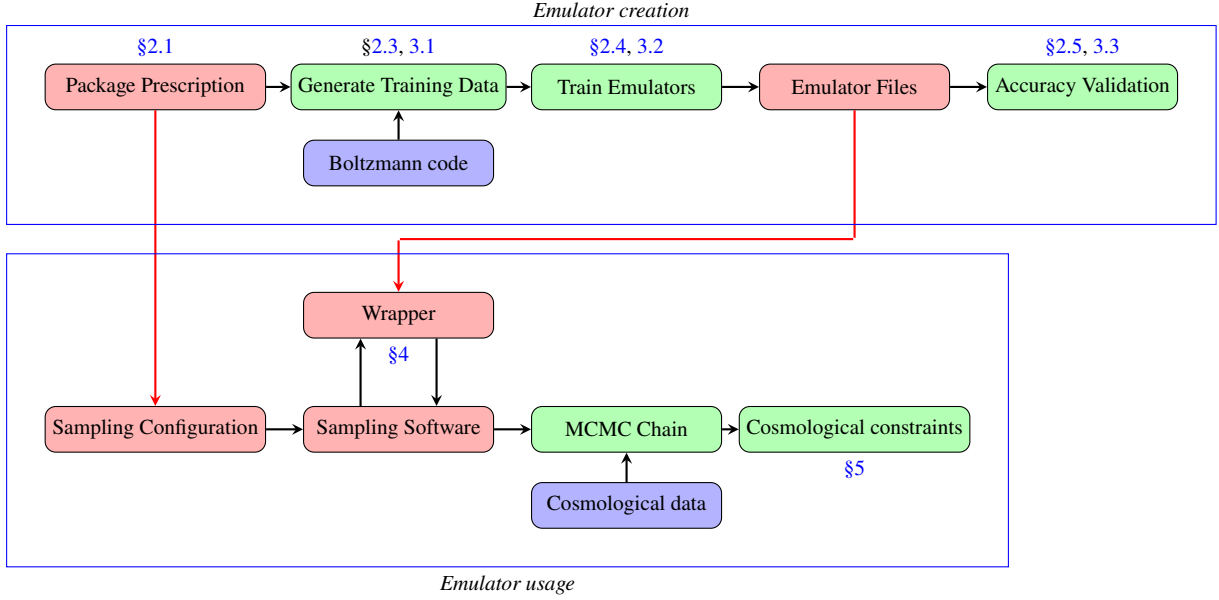

\par In this paper we address the need of standardisation and maintenance of cosmological emulators by devising and releasing a framework which allows one to generate, re-use, and deploy emulators within the major infrastructure tools used by the cosmological community. Our work builds on, and expands, the initial \texttt{CosmoPower}~\citep{CosmoPower2021} software\footnote{\url{https://alessiospuriomancini.github.io/cosmopower/}} and on the development of Stage-IV emulators started in ~\cite{Emulators2023}. We make use of \cosmopower because of its wide range of existing applications \citep{ManciniSpurio:2021jvx,Burger:2022lwh,Heydenreich:2022lsa,Linke:2022xnl,SPT-3G:2022hvq,SpurioMancini:2023mpt,Moretti:2023drg,Reeves:2023kjx,Burger_2024,Farren:2023yna,Carrion:2024itc,Giardiello:2024uzz,Qu:2024lpx}, but the type of packaging and interfaces applied here could also be used with other emulators of Einstein-Boltzmann codes \citep[e.g.][]{Arico:2021izc,Mootoovaloo:2021rot,Nygaard:2022wri,Bonici:2023xjk,Mauland:2023pjt} or other parts of astrophysical forward models. 

\par In ~\cite{Emulators2023}, some of us presented high-accuracy emulators for \class\footnote{Similar emulators for SPT CMB analyses were presented in \cite{SPT-3G:2022hvq}.}. These emulators are capable of reproducing the CMB primary and lensing power spectrum to precision levels $< 10\%$ of the statistical error bars expected from Stage-IV CMB analyses. \cite{Emulators2023} also released emulators for both the linear and non-linear matter power spectrum, as well as background-evolving quantities -- validated using DES-Y1 and BAO analysis likelihoods. In this manuscript, as well as building the framework for community use of these emulators, we present an equivalent suite for \camb~ that are accurate enough for Stage-IV analysis and beyond, demonstrating the emulators for cosmic-variance-limited datasets. We release a full software suite for \python that allows easy creation, testing, and usage of \cosmopower emulators, alongside extensive documentation and example notebooks. This allows the use of our \camb and \class emulators, as well as any future extensions or equivalents, within \cobaya and \cosmosis, which are some of the most commonly used frameworks for Bayesian inference in cosmology.

\par A schematic summary of the new aspects introduced in this paper and how they map into different sections is presented in~\cref{fig:scheme}. More specifically: 
\begin{itemize}[left=3pt,topsep=0pt]
\item In \cref{sec:emulators} we include the details of the Einstein-Boltzmann emulators presented in this work: our models considered, parameter ranges, emulated observables, network structure, and training parameters. We also present the accuracy of these emulators in recovering power spectra.
\item In \cref{sec:packaging_specification} we give an overview of the packaging scheme and \python interface we have developed for Machine Learning emulators. We give details of the specification of pre-trained emulators, and how these are exposed to the software. We also provide examples and guidance for others to create emulators using this framework.
\item In \cref{sec:wrapper-description} we present our wrappers for the \cosmosis and \cobaya sampling software with a brief user guide.
\item In \cref{sec:recover-cosmology} we use these wrappers to run Monte Carlo posterior estimation chains which shows our emulators recover cosmology at the observable level well within the forecast noise ranges of Stage-IV experiments.
\item In \cref{sec:conclusion} we summarise and conclude.
\end{itemize}

\section{Emulators} \label{sec:emulators}
\par In this section we describe the details of our emulators: what is emulated and with which inputs, how an emulation is performed and how the emulators are validated. This serves both as a full description of the emulators released with this manuscript and as guidelines on the creation of new emulators packaged and usable in the same way (e.g. for extended cosmological models). By \emph{emulator} we mean a `black box' code which is capable of ingesting a set of cosmological parameters $\vec{\theta}$ and outputting a set of predictions for the summary statistics of a set of observables $\lbrace \vec{d}_1(\vec{\theta}), \vec{d}_2(\vec{\theta}), \ldots, \vec{d}_N(\vec{\theta}) \rbrace$ which are indistinguishable (within a given tolerance) from the set which would have been produced by a code which explicitly implements numerical models of the physics relating the $\vec{d}$ and $\vec{\theta}$. As the emulation works effectively as an interpolation of the quantities $\vec{d}$ between known points, we rely on the fact that the $\vec{d}$  vary smoothly with respect to the input parameters.

\subsection{Emulated Quantities} \label{subsec:emulated-quantities}

\par In \cref{tab:quantity-ranges} we show the full list of quantities output by the Einstein-Boltzmann code (\camb v.1.5.0) which we focus on emulating in this work. As output observables we generate the CMB temperature, polarisation and lensing potential angular power spectra; linear and non-linear matter power spectra (and their ratio); and a limited set of background expansion and derived perturbation quantities also output by the Einstein-Boltzmann code.

We compute the CMB angular power spectra in the multipole range $2 \le \ell \le 10000$ in each of TT, TE, EE, and BB combinations for different cosmological models. In the basic configurations, we use as inputs for our emulators the six cosmological parameters of the standard $\Lambda$CDM model: the baryon density $\Omega_b h^2$, the dark matter density $\Omega_c h^2$, the amplitude and spectral index of scalar perturbations $\ln(10^{10} A_s)$ and $n_s$, the optical depth to reionization $\tau_\mathrm{reio}$, and the Hubble constant $H_0$ in units of km/s/Mpc. We add additional model parameters to these for separate emulators for $\Lambda$CDM extension models as explained below. When not explicitly varied, neutrinos are described by fixing $N_\mathrm{eff} = 3.044$, corresponding to the contribution from the three Standard Model neutrino species, with one of them carrying a total $0.06$ eV mass.

We also emulate the CMB lensing potential $\phi\phi$ power spectrum in the same multipole range. For this we use the same parameter inputs except for the optical depth to reionization, given that the lensing potential power spectrum does not depend on it.

\par For the matter power spectrum $P(k,z)$, we compute the linear matter power spectrum $P_\mathrm{lin}(k)$ for five input parameters: $\Omega_b h^2$, $\Omega_c h^2$, $\ln(10^{10} A_s)$, $n_s$, and $H_0$, plus again the extra parameters for the extension models. For all matter power spectra we also treat the redshift $z$ as an input parameter, resulting in an emulator function which acts as $P_\mathrm{lin}(k, \vec{\theta})$, where $\vec{\theta}$ includes the redshift. For the non-linear matter power spectrum, we  emulate both the $P_\mathrm{NL}(k)$ spectrum itself and the non-linear boost $P_\mathrm{NL} / P_\mathrm{lin} (k) - 1$. For the emulators included in this paper, we emulate the 2020 version of HMCode described in~\cite{Mead2021}. We sample the wavenumber $k$ at 500 points in the range $10^{-4} \le k \le 50 \, \mathrm{Mpc}^{-1}$ with logarithmic spacing. Note that we compute $P(k)$ up to $k = 100 \, \mathrm{Mpc}^{-1}$ for improved accuracy.

\par For background evolution quantities, we use redshift in the range $0 \le z \le 20$, sampled at 5000 equally-spaced points, as the modes along which we evaluate the redshift-evolution of the Hubble parameter $H(z)$, the angular diameter distance $D_A(z)$, and the clustering $\sigma_8(z)$ for the five input parameters $\Omega_b h^2$, $\Omega_c h^2$, $\ln(10^{10} A_s)$, $n_s$, and $H_0$, plus the extension model parameters where relevant. Adding these background quantities to our emulator packages allows for additional cosmological constraints from e.g., BAO measurements. Additional background quantities, such as $f \sigma_8(z) \equiv -(1+z) \mathrm{d} \sigma_8 / \mathrm{d}z$, can also be easily computed from these quantities with minimal overhead or loss of accuracy. \\

\par We also compute ten derived parameters, namely:

\begin{enumerate}[leftmargin=*]
    \item The angular acoustic scale $\theta_*$ at the surface of last scattering;
    \item The matter clustering parameter $\sigma_8$;
    \item The primordial helium fraction $Y_\mathrm{He}$;
    \item The redshift $z_\mathrm{reio}$ of reionization, defined as the midpoint of reionization described by a simple hyperbolic tangent;
    \item The optical depth $\tau_{r,\mathrm{end}}$ at the end of recombination;
    \item The redshift $z_*$ at the surface of last scattering;
    \item The sound horizon scale $r_*$ at the surface of last scattering;
    \item The redshift $z_d$ at the baryon drag epoch;
    \item The sound horizon scale $r_d$ at the baryon drag epoch;
    \item The effective number of relativistic species $N_\mathrm{eff}$.
\end{enumerate}

\begin{table}
    \centering
    \begin{tabular}{c c l}
        \toprule
        \textbf{Quantity} & \textbf{Range} & \textbf{Emulator} \\
        \midrule
        $C_\ell^{TT}$ & $2 \le \ell \le 10000$ & NN of $\log$-spectra \\
        $C_\ell^{TE}$ & $2 \le \ell \le 10000$ & NN+PCA of spectra \\
        $C_\ell^{EE}$ & $2 \le \ell \le 10000$ & NN of $\log$-spectra \\
        $C_\ell^{BB}$ & $2 \le \ell \le 10000$ & NN of $\log$-spectra \\
        $C_\ell^{\phi\phi}$ & $2 \le \ell \le 10000$ & NN+PCA of $\log$-spectra \\
        $P_\mathrm{lin}(k,z)$ & $10^{-4} \le k \le 50$ & NN of $\log$-spectra \\
        $P_\mathrm{NL}(k,z)$ & $10^{-4} \le k \le 50$ & NN of $\log$-spectra \\
        $P_\mathrm{NL} / P_\mathrm{lin}(k,z)$ & $10^{-4} \le k \le 50$ & NN of spectra ratio \\
        $H(z)$ & $0 \le z \le 20$ & NN of evolution \\
        $\sigma_8(z)$ & $0 \le z \le 20$ & NN of evolution \\
        $D_A(z)$ & $0 \le z \le 20$ & NN of evolution \\
        \hspace{-0.2cm}\emph{derived parameters} & -- & NN of value of derived parameters \\
        \bottomrule
    \end{tabular}
    \caption{Emulated quantities, ranges of scales covered and type of emulator employed for each of them.}
    \label{tab:quantity-ranges}
\end{table}

\par It is common to use $\theta_*$, the angular scale when optical depth is unity, or the approximate parameter $\theta_\mathrm{MC}$, as a sampled parameter in MCMC analyses of CMB data due to its lower level of covariance with other parameters than $H_0$\footnote{see note at \url{https://cosmologist.info/cosmomc/readme.html}}. As we also noted in~\cite{Emulators2023} however, \camb and \class use different points at which to evaluate the angular scale (with \class defining $\theta_s$ as the angular scale at maximum visibility, which is close to but not the same as $\theta_*$, which is used in \camb). To maintain cross-compatibility between our emulators, and to remain consistent with our earlier work, we therefore use $H_0$ as an input, and not $\theta_*$. Including these derived parameters as emulators allow us to recover the posterior distributions on these quantities, either directly storing their computed values while sampling the chain, or afterwards by post-processing a converged MCMC chain.

\subsection{Cosmological Models} \label{subsec:models-extensions}
\par We provide emulators for the $\Lambda$CDM model with parameters $\lbrace \Omega_b h^2, \Omega_c h^2, \ln(10^{10} A_s), n_s, H_0, \tau_{\rm reio} \rbrace$ defined above as well as the following four extended models:
\begin{enumerate}[leftmargin=*]
    \item $\Lambda$CDM+$N_\mathrm{eff}$: varying the effective number of relativistic species $N_\mathrm{eff}$;
    \item $\Lambda$CDM$+\Sigma m_\nu$: varying the sum of neutrino masses $\Sigma m_\nu$;
    \item $\Lambda$CDM$+N_\mathrm{eff} \Sigma m_\nu$: varying both the number and mass sum of neutrinos;
    \item $\Lambda$CDM+$w_0 w_a$: varying the dark energy equation of state described with two parameters $w_0$ and $w_a$.
\end{enumerate}

\par Each of these four extension models is emulated separately, with the extension parameters used as additional inputs. We chose to emulate $+N_\mathrm{eff}$ and $+\Sigma m_\nu$ separately, and the combination $+N_{\rm eff}+\Sigma m_\nu$ to explore the relation between model complexity and emulator accuracy. While, as we show later, our emulator for the full combination $+N_{\rm eff}+\Sigma m_\nu$ is accurate enough for cosmological analysis, we release the single parameter-extension model emulators as they offer greater accuracy over the higher-dimensional models.

\par For the non-linear $P_{NL}(k,z)$ and non-linear boost $P_{NL} / P_L(k,z) - 1$ emulators, we include the baryonic feedback parameters $A_b$, $\eta_b$, and $\log T_\mathrm{AGN}$ that appear in HMCode~\citep{Mead2021} and are otherwise fixed at their default \camb values in the other emulators. For the remaining model choices, we set a primordial helium fraction set from BBN consistency using PRIMAT~\citep{PitrouEtal2018}, recombination from the \texttt{CosmoRec} code~\citep{Chluba_2010a,Chluba_2010b}, and reionization modeled with a simple hyperbolic tangent with a redshift width $\Delta z = 0.5$. Most of these options are the default settings in \camb. We only changed the recombination code to \texttt{CosmoRec}, whereas the \camb default is to use the older \texttt{RECFAST} code.

\subsection{Training Data} \label{subsec:training-data}

\par \emph{Training} of emulators involves creating a set of output data $\vec{d}$ at a finite sample of known parameter values $\vec{\theta}$ using the code to be emulated (i.e. the Einstein-Boltzmann code here). These data will then subsequently be used in \cref{subsec:network-training} for the neural network to learn an approximate (but high accuracy) mapping between input and output. Training data must be generated at a high enough resolution in the input parameters that we can smoothly interpolate between outputs. The training data only need to be generated once, to train the emulator, and do not need to be generated using the computationally intensive numerical code again in any subsequent inference.

\begin{figure}
    \centering
    \includegraphics[width=0.45\textwidth]{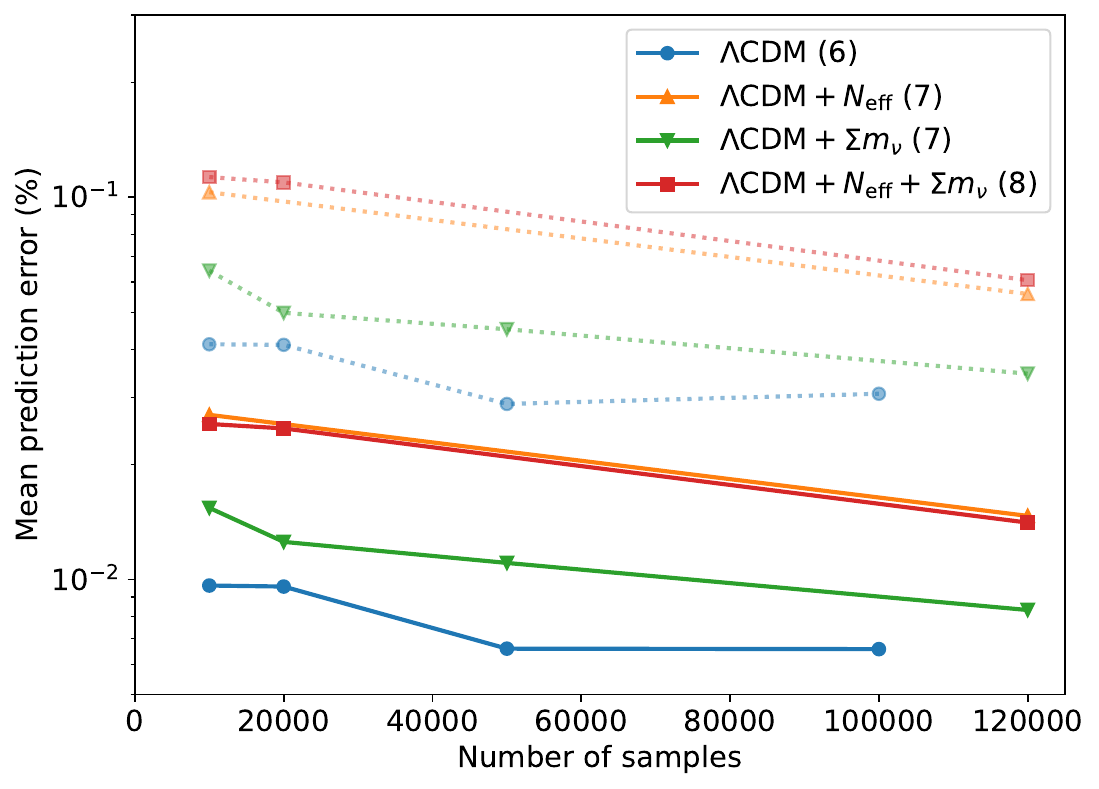}
    \caption{An overview of the accuracy reached by a trained $C_\mathrm{\ell}^{TT}$ emulator given the number of training spectra used to train the emulator, for an increasing number of input parameters. The solid coloured lines and point represent the 68\% error of a $C_\ell^{TT}$ emulator trained with $N_S$ samples (the dotted shaded lines and points show the similar behaviour observed at 99\%), averaged over the entire $\ell$-range. We show the full-size emulators generated with $N_S=100000$ for $\Lambda$CDM and $N_S=120000$ for extended models, as well as emulators with a smaller training set to show how accuracy scales with $N_S$ and input parameters. We train emulators for $\Lambda$CDM (6 parameters, \textbf{blue}), $+\Sigma m_\nu$ (7 parameters, \textbf{orange}), and $+N_\mathrm{eff}+\Sigma m_\nu$ (8 parameters, \textbf{green}), each on a random smaller subset of the full training dataset, scaling the training batch size proportional to the size of the subset. We show how the mean emulation error decreases as the number of training spectra increases, and increases as we increase the complexity of the parameter space. We note, however, that scaling of emulators accuracy with number of input parameters is non linear, the nature and impact on the emulated quantity of the specific parameter will matter for this behaviour.}
    \label{fig:accuracy-vs-number-of-spectra}
\end{figure}

Following \cite{CosmoPower2021} and \cite{Emulators2023}, we generate $N_S = 10^5$ sets of output spectra as training data, of which $20\%$ will be used for validating the network accuracy, and the rest for training. Our parameter space is shown in~\cref{tab:parameter-ranges}. We employ Latin Hypercube (LHC) sampling for ensuring our parameter space is evenly sampled. For extended models, we choose to generate slightly more spectra at $N_S = 1.2 \times 10^5$, to compensate for the expanded parameter space. To demonstrate the need for this and to provide some guidance on how to select $N_S$, we show a comparison of the mean prediction error versus the size of the training dataset in~\cref{fig:accuracy-vs-number-of-spectra}, for a varying number of input parameters. The figure shows that there is not a simple linear scaling with the number of parameters. Although increasing the number of parameters always requires a larger training set to reach the desired target accuracy, the physical nature and range of variation of the specific additional parameter will impact the results. For example, if we extend \lcdm varying $N_\mathrm{eff}$ or $\Sigma m_\nu$, we observe different behaviours, even if in both cases it is only one additional input parameter (7 input parameters compared to 6 for \lcdm). We explain this by noting that cosmological observables have different responses to different parameters, according to the physics signature they are tracking. For example, the CMB $C_\ell^{TT}$ spectrum will exhibit a strong dependence on $N_\mathrm{eff}$ -- changing both the peak position and amplitude at all scales, but less so on $\Sigma m_\nu$ which will primarily appear at scales dominated by lensing. Hence in~\cref{fig:accuracy-vs-number-of-spectra} the $\Lambda$CDM$+N_\mathrm{eff}$ case requires more training than $\Lambda$CDM$+\Sigma m_\nu$. When we expand further the model to $\Lambda$CDM$+N_\mathrm{eff}\Sigma m_\nu$ (8 input parameters compared to 6 for \lcdm), we observe a very similar behaviour to the 7-parameter case \lcdm+$N_\mathrm{eff}$, because we have already covered most of the strongly-varying training region. We conclude that to achieve the desired convergence of the emulators, the user will need to monitor the behaviour of their specific model and perform some exploratory studies of how the emulators depend on the model parameters. 

\begin{table}
    \centering
    \begin{tabular}{c c c}
        \toprule
        \textbf{Parameter} & \textbf{Range} & \textbf{Default Value}\\
        \midrule
        $\Omega_b h^2$ & $[0.015, 0.03]$ & --\\
        $\Omega_c h^2$ & $[0.09, 0.15]$ & -- \\
        $\ln(10^{10} A_s)$ & $[2.5, 3.5]$ & -- \\
        $n_s$ & $[0.85, 1.05]$ & -- \\
        $\tau_\mathrm{reio}$ & $[0.02, 0.20]$ & -- \\
        $H_0$ [km/s/Mpc] & $[40, 100]$ & -- \\
        \midrule
        $z_\mathrm{pk}$ & $[0, 5]$ & -- \\
        $A_b$ & $[2, 4]$ & 3.13 \\
        $\eta_b$ & $[0.5, 1.0]$ & 0.603 \\
        $\log T_\mathrm{AGN}$ & $[7.3, 8.3]$ & 7.8 \\
        \midrule
        $N_\mathrm{eff}$ & $[1.5, 5.5]$ &  3.044 \\
        $\Sigma m_\nu$ [eV] & $[0, 0.5]$ & 0.06 \\
        $w_0$ & $[-2, 0]$ & -1.0 \\
        $w_a$ & $[-2, 2]$ & 0.0 \\
        \bottomrule
    \end{tabular}
    \caption{Table of parameter ranges over which we trained our emulators. Compare this with the textual specification in~\cref{snip:parameters}. The \textbf{top} section of the table refers to the background cosmology parameters used in almost all emulators. The \textbf{middle} section of the table contains the redshift and baryonic feedback parameters used only in the $P(k)$ emulators, with their default values from \camb used in the CMB and background evolution emulators. The \textbf{bottom} section of the table shows the ranges of the single-/two-parameter extension model emulators, and their default values taken in the base \lcdm case. Each emulator takes in the first six parameters, and one or two extension parameters, with the exception for $C_\ell^{\phi\phi}$, and background quantities, which do not rely on $\tau_\mathrm{reio}$.}
    \label{tab:parameter-ranges}
\end{table}

\par To meet the requirements for Stage-IV analyses, we use the \camb accuracy settings suggested by~\cite{McCarthy2022,Hill2022} as adequate for convergence of the likelihood value obtained from data with this level of precision, summarised in~\cref{snip:accuracy}.

\begin{figure}
    \begin{lstlisting}[language=yaml]
lmax: 10000
kmax: 10.0
k_per_logint: 130
nonlinear: True
lens_potential_accuracy: 8
lens_margin: 2050
lAccuracyBoost: 1.2
min_l_logl_sampling: 6000
DoLateRadTruncation: False
recombination_model: CosmoRec
    \end{lstlisting}
    \caption{Accuracy settings for \camb, based on the settings earlier suggested in \protect\cite{McCarthy2022,Hill2022}. For an example of a full yaml file, see~\cref{sec:lcdm-prescription}.}
    \label{snip:accuracy}
\end{figure}

\par We iterate over each of the $N_S$ samples in our LHC, computing the CMB, lensing, and matter power spectra, as well as background quantities, and derived parameters with \camb (see~\cref{tab:quantity-ranges} for a summary of the outputs and their ranges), and store the results in a structured data file containing appropriate metadata (see~\cref{sec:file-structure}). Because of our choices of parameter limits as a hypercube, a small fraction ($\ll 1\%$) of our samples are in unphysical parts of parameter space and can cause issues in computations from \camb. Because this number is small, these samples are simply discarded and ignored for future processing.

\subsection{Network Design and Training} \label{subsec:network-training}

\par Following~\cite{CosmoPower2021}, we implement the emulators as dense neural networks, with four hidden layers of 512 neurons each. Each emulator takes the normalised parameters as input, and maps it to normalised spectra. We use the activation function from~\cite{CosmoPower2021}:

\begin{equation}
    f(\vec{x}) = \left[ \vec{\gamma} + \left( 1 + e^{-\vec{\beta} \odot \vec{x}} \right)^{-1} \odot (1 - \vec{\gamma}) \right] \odot \vec{x} ,
\end{equation}

\par where $\odot$ is the element-wise product. For the optimizer, we re-use the Adam optimizer. The input and output quantities are normalised with respect to mean and standard deviations of the respective ranges. For most quantities, as detailed in~\cref{tab:quantity-ranges} we emulate the logarithm of the spectrum, as the high dynamic range of these values makes it easier for the emulator to reconstruct the log-values. We employ the same method for the background quantities $H(z)$, $\sigma_8(z)$, and $D_A(z)$, where we reconstruct the logarithm of the redshift evolution. 

\begin{figure*}
    \centering
    \begin{subfigure}{0.4\textwidth}
        \includegraphics[width=\textwidth]{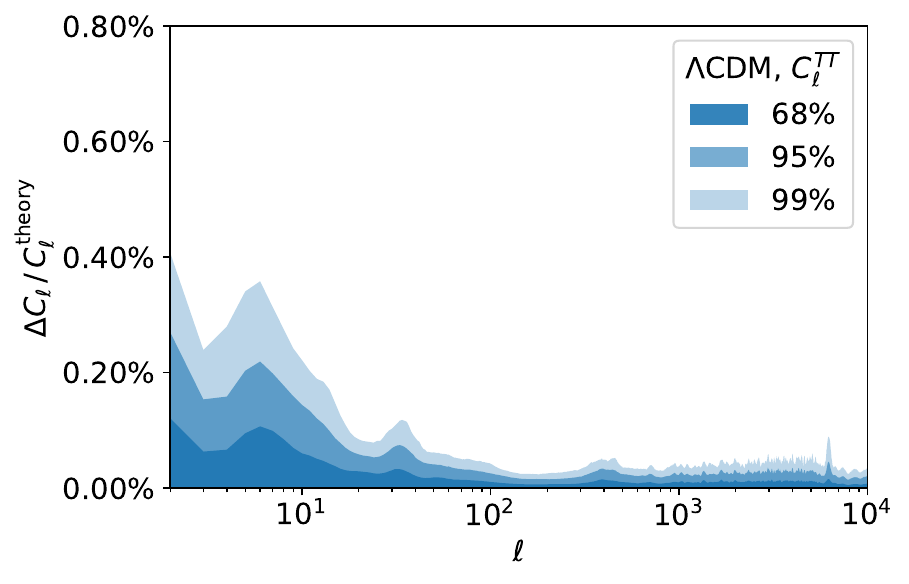}
         \vspace{-0.6cm}\caption{TT}
        \label{fig:lcdm-frac-cl-tt}
    \end{subfigure}
    \begin{subfigure}{0.4\textwidth}
        \includegraphics[width=\textwidth]{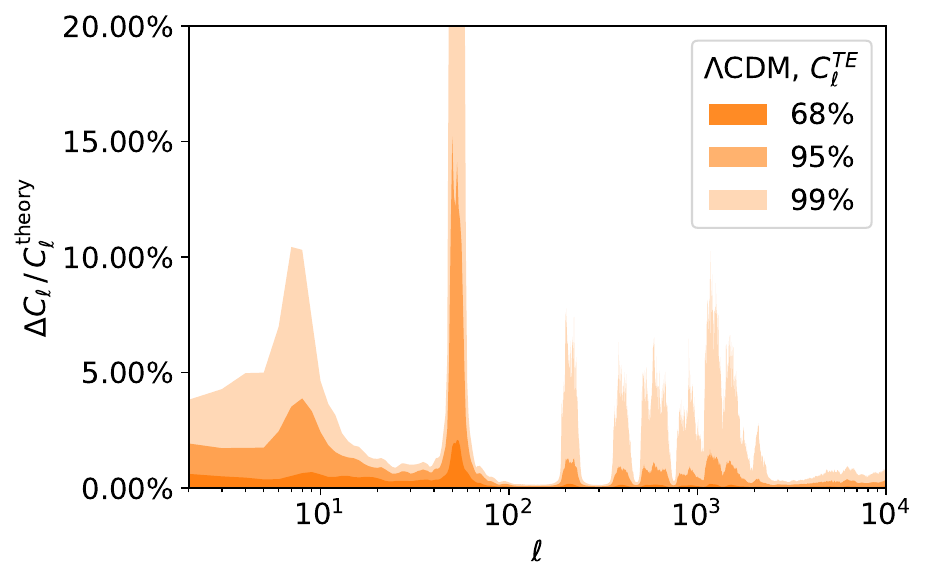}
        \vspace{-0.6cm}\caption{TE}
        \label{fig:lcdm-frac-cl-te}
    \end{subfigure}
    \begin{subfigure}{0.4\textwidth}
        \includegraphics[width=\textwidth]{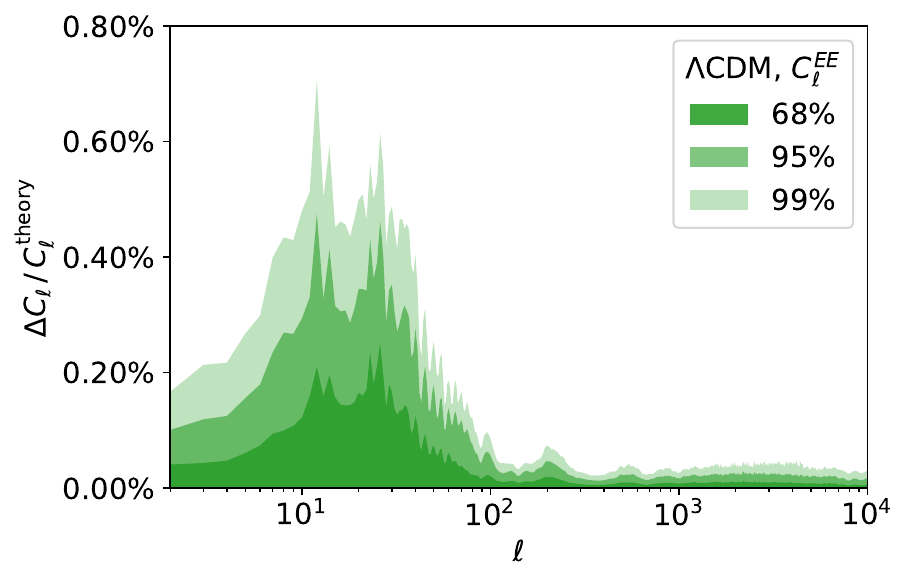}
         \vspace{-0.6cm}\caption{EE}
        \label{fig:lcdm-frac-cl-ee}
    \end{subfigure}
    \begin{subfigure}{0.4\textwidth}
        \includegraphics[width=\textwidth]{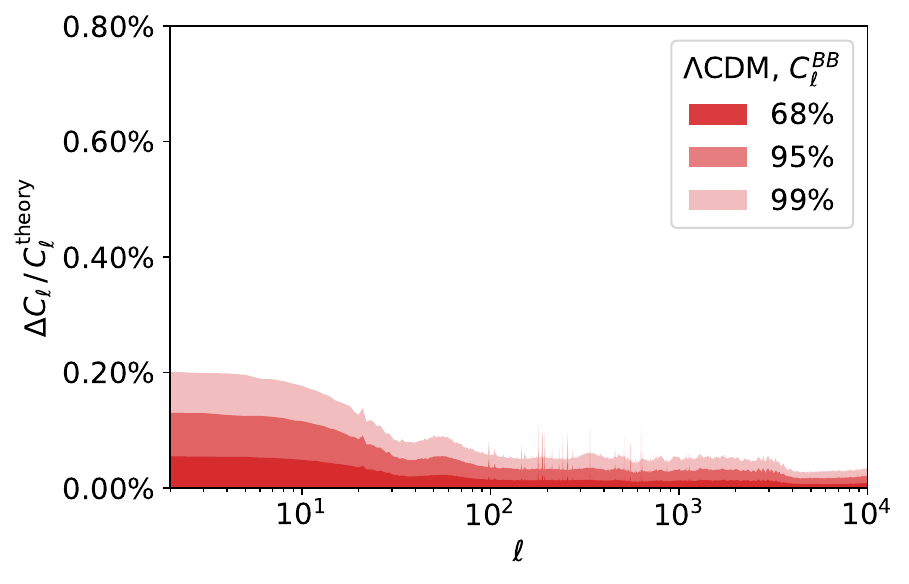}
         \vspace{-0.6cm}\caption{BB}
        \label{fig:lcdm-frac-cl-bb}
    \end{subfigure}
    \begin{subfigure}{0.4\textwidth}
        \includegraphics[width=\textwidth]{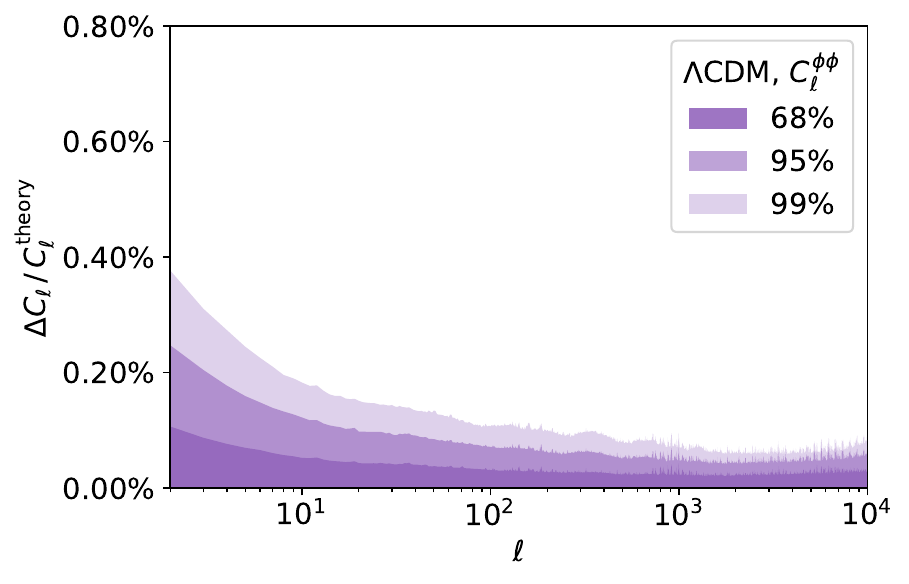}
         \vspace{-0.6cm}\caption{$\phi\phi$}
        \label{fig:lcdm-frac-cl-pp}
    \end{subfigure}
    \begin{subfigure}{0.39\textwidth}
        \includegraphics[width=\textwidth]{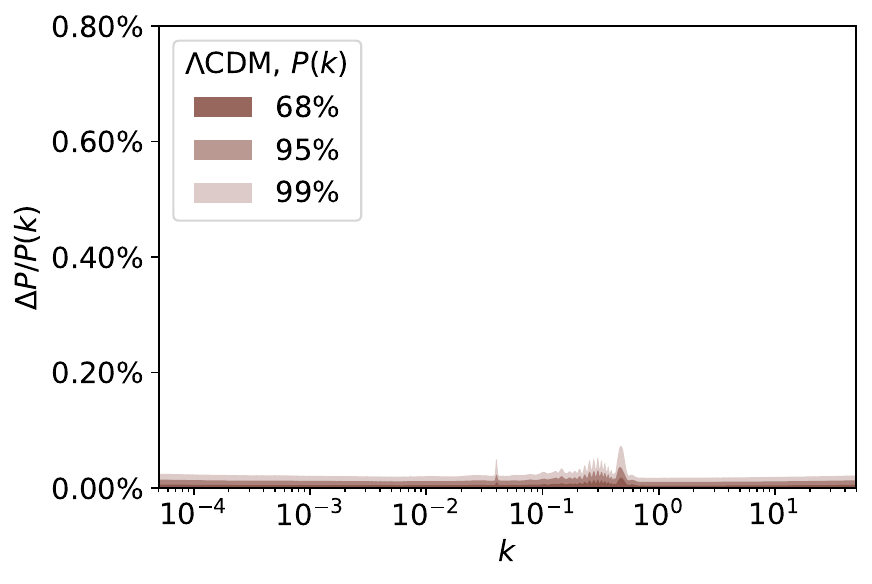}
         \vspace{-0.6cm}\caption{$P(k)$}
        \label{fig:lcdm-frac-cl-pk}
    \end{subfigure}
    \caption{A validation graph generated from our trained networks for $\Lambda$CDM. We show the error in the reconstructed CMB power spectrum in $C_\ell^{TT}$ (blue, top-left), $C_\ell^{TE}$ (orange, top-right), $C_\ell^{EE}$ (green, centre-left), $C_\ell^{BB}$ (red, centre-left), $C_\ell^{\phi\phi}$ (purple, bottom-left), and linear $P_\mathrm{lin}(k)$ (brown, bottom-right) relative to the \camb theory curve. The bands show the $68/95/99\%$ contours (from darkest to lightest shades). Note the different scale for TE, for which errors get blown up due to the zero-crossings of the input power spectrum.}
    \label{fig:deltas}
\end{figure*}

\par For the $C_\ell^{TE}$ emulator, the resulting raw spectra include zero-crossings which make emulating the log-spectra impossible. Because the unscaled spectra still contain a high dynamic range in values, we follow~\cite{CosmoPower2021} in first decomposing the spectra with a Principal Component Analysis (PCA) and then subsequently emulating the sets of PCs. Similar to before, we decompose the $C_\ell^{TE}$ spectra into 512 PCs. Even though they remain completely positive, we also decompose the $C_\ell^{\phi\phi}$ spectra into 64 PCs. We find that this is more effective at emulating the $\phi\phi$ spectra, which we explain with the reduced dimensionality of the information contained in the $\phi\phi$ spectra. We introduce the procedure of constructing scree plots, showing the eigenvalues associated with each PC in the decomposition, to identify the ``elbow'' at which higher PC numbers no longer carry significant weight and can be discarded. For more details regarding this and for guidance on decisions regarding PCA see \cref{sec:pca-appendix}. With this setup, our emulator design for the CMB spectra remains fully consistent with the original emulators from~\cite{CosmoPower2021}. 

\par For the matter power spectra $P_\mathrm{lin}(k,z_\mathrm{pk})$ and $P_\mathrm{NL}(k,z_\mathrm{pk})$, we choose to emulate $\log P_\mathrm{lin}(k,z_\mathrm{pk})$ and the non-linear boost $P_\mathrm{NL} / P_\mathrm{lin}(k,z_\mathrm{pk}) - 1$ for best performance. These quantities are functions of two parameters, the wavenumber $k$ and redshift $z_\mathrm{pk}$. Similar to previous emulators we have developed, we use $k$ as the one-dimensional grid along which we sample our spectra, and use $z_\mathrm{pk}$ as an additional input for our $P(k)$ emulators.

\par The time it takes to train an emulator depends on many factors, including the size of the dataset, the number of inputs and outputs of the network, the hardware performance, as well as some inherently stochastic factors in the training process. At $10^5$ training samples for a network, we find it takes $O(1 h)$ to train a $C_\ell$ network on a GPU. If no GPU hardware or the required software is available, then the emulators can alternatively be trained on a CPU, which for the same case still only takes $O(10 h)$ to perform.

\subsection{Accuracy of Emulated Observables} \label{subsec:accuracy-validation}
To assess the accuracy of our emulators we perform a number of comparisons between the observables emulated and those calculated directly with \camb. This allows us to understand if we have reached the theoretical calculation accuracy required for Stage-IV analyses. This functionality is now fully built into our released software as described later in~\cref{subsec:accuracy-plotting}.

\begin{figure}
    \centering
    \includegraphics[width=\columnwidth]{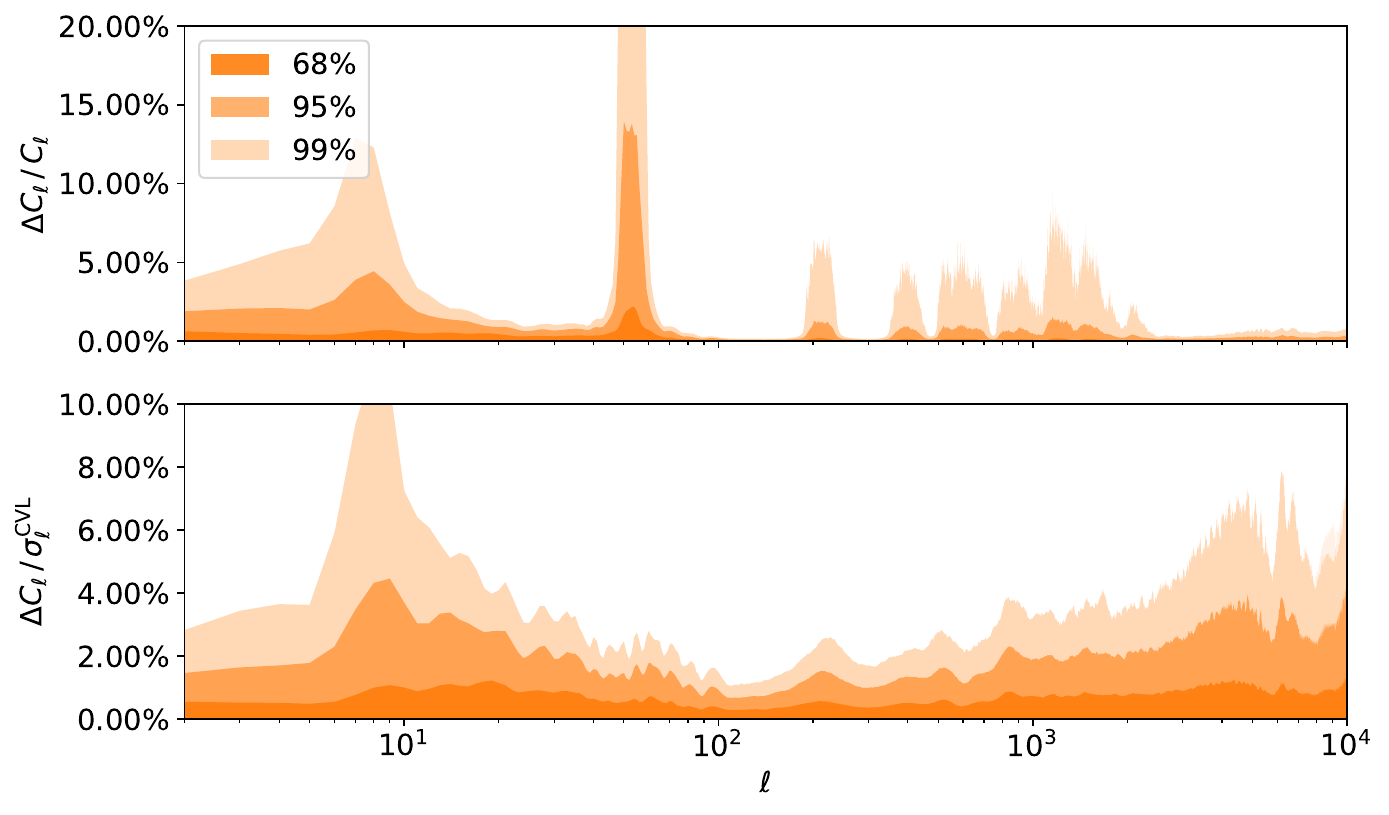}
    \caption{A direct comparison of the error in the $C_\ell^{TE}$ emulator as measured in fractional error with respect to the training spectrum (\textbf{top} and as in \cref{fig:deltas}), and as relative error with respect to a cosmic variance-limited noise curve (\textbf{bottom} and as in \cref{fig:lcdm-cmb-cl-validation} with more details in ~\cref{eq:cmb-noise-curve}). The peaks in the top figure are due to the zero-crossings of the $C_\ell^{TE}$ power spectrum, which ``blow up'' any errors in the emulator. Using a cosmic variance limit noise curve provides a more realistic error measure, as shown in the bottom figure, where the inclusion of the $C_\ell^{TT}$ and $C_\ell^{EE}$ terms in the error wash out these zeroes and provide a more reasonable assessment for the error.}
    \label{fig:clte-delta-sigma}
\end{figure}

In \cref{fig:deltas} we report the difference between direct \camb outputs and emulated observables, showing contours corresponding to the fraction of our training spectra (across the full parameter space) which lie within a given level of agreement with the emulated values. All the CMB spectra reach sub-percent accuracy (note that the TE higher values are numerical artefacts due to diving for a signal crossing zero, see~\cref{fig:clte-delta-sigma} for more details); the matter power spectrum is accurate at the few percent level relative to the \camb prediction for very large range of wave numbers.

\par For the CMB observables, as done in previous works we can also compute the difference relative to (or `in units of') a specific experiment's sensitivity which tracks the noise for each observable $N^{XY}_\ell$ with

\begin{eqnarray} \label{eq:cmb-noise-curve}
     &\sigma^{XY}_\ell & =  \\
     && \hspace{-1cm}\sqrt{\frac{1}{f_\mathrm{sky} (2 \ell + 1)} \left[ (C_\ell^{XX} + N_\ell^{XX})(C_\ell^{YY} + N_\ell^{YY}) + (C_\ell^{XY} + N_\ell^{XY})^2 \right]},\nonumber 
\end{eqnarray}
where, for the cosmic variance limit, $f_\mathrm{sky} = 1$ and $N_\ell^{XY} = 0$ for all $XY$.

\begin{figure*}
    \centering
    \begin{subfigure}{0.4\textwidth}
        \includegraphics[width=\textwidth]{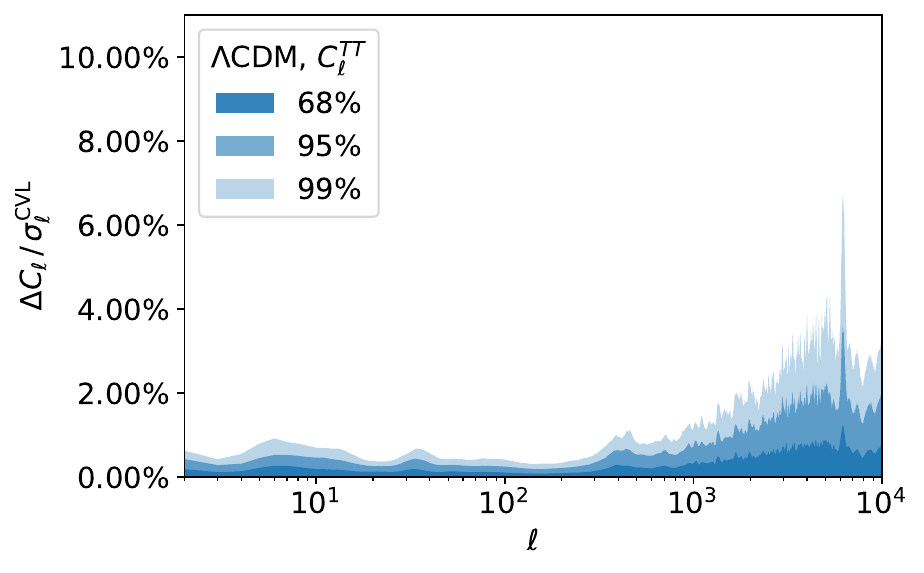}
        \vspace{-0.6cm} \caption{TT}
        \label{fig:lcdm-cvl-cl-tt}
    \end{subfigure}
    \begin{subfigure}{0.4\textwidth}
        \includegraphics[width=\textwidth]{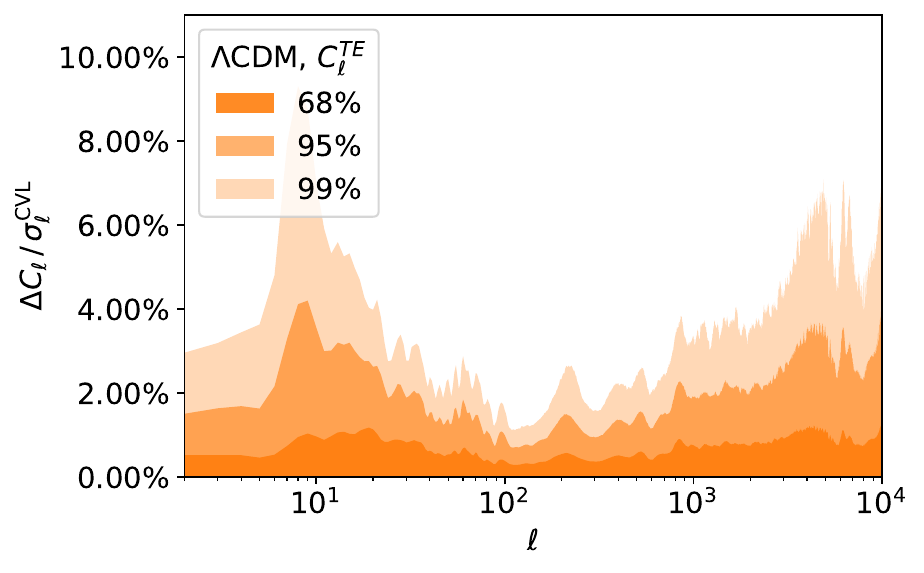}
        \vspace{-0.6cm} \caption{TE}
        \label{fig:lcdm-cvl-cl-te}
    \end{subfigure}
    \begin{subfigure}{0.4\textwidth}
        \includegraphics[width=\textwidth]{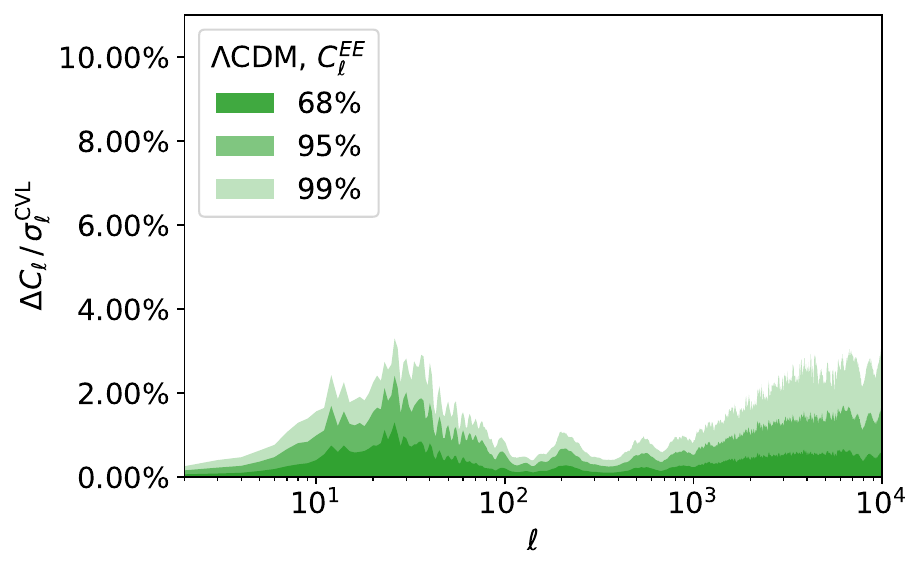}
        \vspace{-0.6cm} \caption{EE}
        \label{fig:lcdm-cvl-cl-ee}
    \end{subfigure}
    \begin{subfigure}{0.4\textwidth}
        \includegraphics[width=\textwidth]{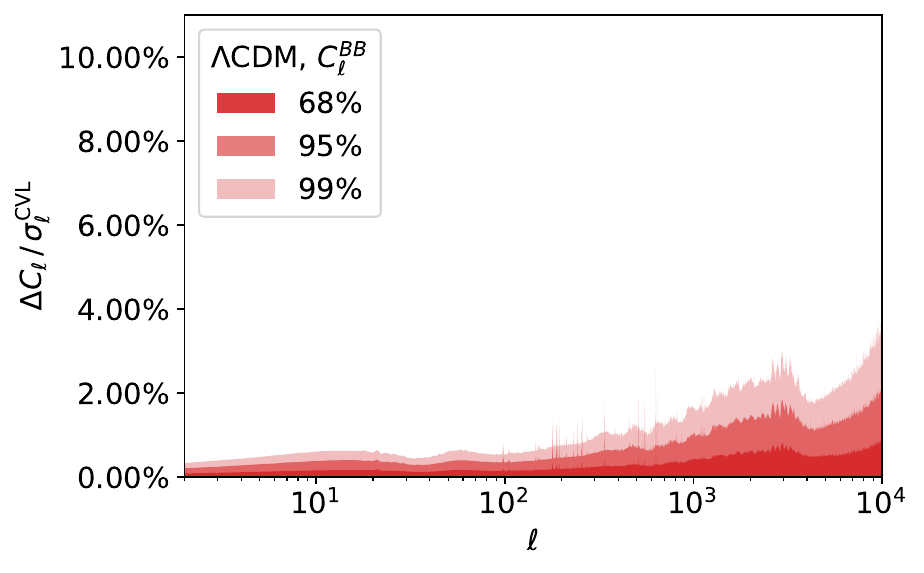}
        \vspace{-0.6cm} \caption{BB}
        \label{fig:lcdm-cvl-cl-bb}
    \end{subfigure}
    \begin{subfigure}{0.4\textwidth}
        \includegraphics[width=\textwidth]{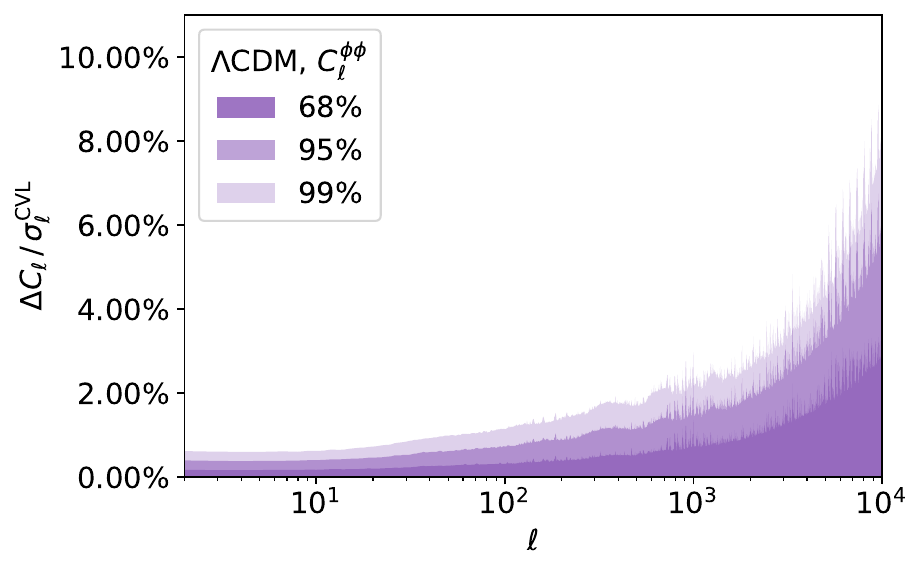}
        \vspace{-0.6cm} \caption{$\phi\phi$}
        \label{fig:lcdm-cvl-cl-pp}
    \end{subfigure}
    \begin{subfigure}{0.38\textwidth}
        \includegraphics[width=\textwidth]{figs/accuracy/lcdm_pk_Pk_lin}
        \vspace{-0.6cm} \caption{$P(k)$}
        \label{fig:lcdm-cvl-cl-Pk}
    \end{subfigure}
    \caption{A validation graph generated from our trained networks for $\Lambda$CDM. We show the recovered CMB power spectrum $C_\ell^{TT}$ (blue, top-left), $C_\ell^{TE}$ (orange, top-right), $C_\ell^{EE}$ (green, center-left), $C_\ell^{BB}$ (red, center-right), $C_\ell^{\phi\phi}$ (purple, bottom-left), and linear $P_\mathrm{lin}(k)$ (brown, bottom-right), with respect to the cosmic variance limit for $C_\ell$'s, and as a fractional difference for $P(k)$. The bands show the $68/95/99\%$ contours (from darkest to lightest shades).}
    \label{fig:lcdm-cmb-cl-validation}
\end{figure*}

\par We show this accuracy of our emulators relative to the cosmic variance-limited experimental noise for \lcdm in~\cref{fig:lcdm-cmb-cl-validation} and extended models are shown in \cref{sec:extension_accuracy} in~\cref{fig:neff-cmb-cl-validation,fig:mnu-cmb-cl-validation,fig:neff-mnu-cmb-cl-validation,fig:w0wa-cmb-cl-validation} (for \lcdm\xspace$+N_\mathrm{eff}$, $+\Sigma m_\nu$, $+N_\mathrm{eff} \Sigma m_\nu$, and $+w_0 w_a$ respectively).

\par All our emulators remain well within $10\%$ of a cosmic variance-limited experimental uncertainty range. The only exception to this is our $w_0 w_a$ emulator (see~\cref{fig:w0wa-cmb-cl-validation}), for which some outliers at small scales in the CMB emulators can reach about $80\%$ of this uncertainty. We attribute this effect to the parameter degeneracy of the model, as well as the complexity of this model and the relatively wide range of parameters we chose. However, in the absence of CMB sensitivity to the mechanics of dark energy, and in the interest of the recent results from DESI~\citep{desicollaboration2024desi}, we are still including this emulator.

\section{Packaging Description} \label{sec:packaging_specification}

\par As part of this release, alongside new emulators we build a packaging prescription for \cosmopower emulators. This prescription is both human- and machine-readable and serves as a description of what the emulator is capable of and its full design specifications. The \cosmopower software package\footnote{\url{https://github.com/alessiospuriomancini/cosmopower}} has been updated to include a full parser for the packaging prescription. 

\par To create and train a new emulator, the packaging prescription is designed to guide both the author and a later user through the process of considering what quantities are emulated, how, and to what accuracy.

\par In this section, we describe the main steps of creating an emulator, namely:
(1) describing the input parameters and output data, and generating the training spectra with the Einstein-Boltzmann code, (2) detailing the specifications of the emulator and the training parameters, and performing the training process, and (3) testing the validation of emulators. We follow the creation of the emulators we specified in~\cref{sec:emulators}, and describe how the packaging prescription of these emulators is setup, as well as alternative options and choices available for the user.

\par We also create and release packaging for the emulators for the \class Einstein-Boltzmann code presented in~\cite{Emulators2023} which also achieve Stage-IV-level accuracy,  consistent with the \camb emulators in this work. With our included packaging, these emulators can likewise be used in the inference frameworks with the same level of convenience and robustness.

\subsection{Generating Training Data} \label{subsec:generate-training}

\par In this subsection, we discuss the required prescription of the input parameters for emulators, and for the output of quantities that are desired to be emulated.

\par As mentioned above, \cosmopower uses LHC sampling, which allows for an evenly spaced grid of sampling points that are sufficiently distributed that the entire parameter space is covered with minimal variation in sampling density. In~\cref{snip:parameters} we show how to specify the LHC grid in the prescription file.

\begin{figure}
    \begin{lstlisting}[language=yaml]
emulated_code:
  name: camb
  version: "1.5.0"
  inputs: [ombh2, omch2, As, ns, H0, tau]
  extra_args:
    <...>

samples:
  Ntraining: 100000
  parameters:
    ombh2: [0.015,0.03]
    omch2: [0.09,0.15]
    # We want to sample on log(10^10 As), but our
    # Boltzmann code takes As as an input.
    logA: [2.5,3.5]
    As: "lambda logA: 1.e-10 * np.exp(logA)"
    tau: [0.02, 0.20]
    ns: [0.85, 1.05]
    H0: [40.0, 100.0]
    # Parameters computed by the Boltzmann code
    thetastar:
    sigma8:
    YHe:
    zrei:
    taurend:
    zstar:
    rstar:
    zdrag:
    rdrag:
    Neff:
    \end{lstlisting}
    \caption{Code snippet for sampling and parameters block, compare this with~\cref{tab:parameter-ranges}. In the example here, we setup the aforementioned six parameters to sample over, add an intermediate parameter $A_s$, and add the nine parameters which are derived directly from the Boltzmann code, in this case \camb. Note that \camb expects the primordial amplitude $A_s$ to be provided, but it is far more common to sample over $\ln(10^{10} A_s)$ instead. By defining the \texttt{logA} parameter and marking the \texttt{As} parameter as a derived parameter from that, we can perfectly accomplish this. At the bottom we show the nine parameters we derive from the Boltzmann code - in this case, they are computed by \camb. It is possible to use any of the parameters defined in this block as an input to the networks, including the parameters derived from the Boltzmann code. The \texttt{extra\_args} block would include any accuracy settings, as seen in~\cref{snip:accuracy}.}
    \label{snip:parameters}
\end{figure}

\par The \texttt{emulated\_code} block of the packaging contains information about the Einstein-Boltzmann code being emulated, in particular the name and version number. If a customized version of a code is used, it is possible to manually specify the import path with the \texttt{boltzmann\_path} keyword. The \texttt{inputs} keyword is the list of named parameters which will be varied as inputs to the Einstein-Boltzmann code. \texttt{extra\_args} contains code parameters which embody any model choices or approximation and accuracy settings.

\par The \texttt{samples} block specifies the \texttt{Ntraining} training spectra to be generated. The packaging prescription recognises four different types of parameters in the \texttt{parameters} block:

\begin{enumerate}[leftmargin=*]
    \item Sampled parameters, these are the parameters that the LHC is created over, and are defined with a minimum-maximum pair for the range over which the LHC is sampled, e.g. \texttt{ombh2: [0.015, 0.03]};
    \item Derived parameters, these are parameters that are trivially derived from other sampled parameters, and are defined with a text string prescribing a python lambda function equation to derive them directly, e.g. \texttt{As: "lambda logA: 1.e-10 * np.exp(logA)"};
    \item Fixed parameters, these are simply defined by writing a single numerical value that the parameter is set to, e.g. \texttt{mnu: 0.06};
    \item Computed parameters, these are parameters that we cannot easily compute ourselves, but the Boltzmann code can, and these are defined by simply leaving an empty tag in the parameter list. These parameters are specified by variable names available to \cosmopower at the spectra generation stage via the python interfaces of the Einstein-Boltzmann codes being emulated, e.g. ``\texttt{YHe: }'' for $Y_{\rm He}$.
\end{enumerate}

\par Any of these types of parameters can be used as an input to a network, and any of the first three types can be used as an input for the Einstein-Boltzmann code. It is for example possible to create an LHC over a range of Hubble parameter $H_0$, while using the angular scale $\theta_*$, as computed by the Einstein-Boltzmann code, as an input for the emulators. 

\par The \texttt{networks} block specifies the neural networks to be created using the training data. It is possible to specify multiple networks, each under a \texttt{quantity} heading, which each have their own set of network properties specified as further blocks and keywords. When creating \cosmopower networks, the current list of quantities to which can emulated is defined and described as follows:
\begin{itemize}[leftmargin=*]
    \item \texttt{Cl/xy}: referring to (lensed) CMB angular power spectra $C^{XY}_\ell$ with $X,Y$ any combination of T/E/B ($C_\ell^{TT}$, $C_\ell^{TE}$, $C_\ell^{EE}$, $C_\ell^{TB}$, $C_\ell^{EB}$, and $C_\ell^{BB}$);
    \item \texttt{Cl/pp}: CMB lensing potential spectrum for $C_\ell^{\phi\phi}$, there are also options available for cross-spectra with primary CMB via \texttt{Cl/pt}, \texttt{Cl/pe}, and \texttt{Cl/pb};
    \item \texttt{Pk/lin} and \texttt{Pk/nonlin}: Matter power spectrum for linear $P_\mathrm{lin}(k,z)$ and non-linear $P_\mathrm{nl}(k,z)$;
    \item \texttt{Pk/nlboost}: The non-linear boost $(P_\mathrm{NL} / P_\mathrm{lin} - 1)(k,z)$ defined as the non-linear boost to the linear matter power spectrum;
    \item \texttt{Hubble}, \texttt{Omegab}, \texttt{Omegac}, \texttt{Omegam}, \texttt{sigma8} and \texttt{DA}: The redshift-evolving quantities $H(z)$, $\Omega_b(z)$, $\Omega_c(z)$, $\Omega_m(z)$, $\sigma_8(z)$, and $D_A(z)$.
\end{itemize}
\par It is also possible to specify \texttt{derived} quantities. This network will automatically use all parameters from the \texttt{parameter} block that are computed by the Einstein-Boltzmann code as outputs. So, when we specify a \texttt{derived} network in our emulators similar to our $C_\ell^{TT}$ emulator, we create an emulator that emulates the computation of the nine quantities mentioned in~\cref{subsec:emulated-quantities} (which are the nine parameters we listed in~\cref{snip:parameters}). 

\par In \cref{snip:networks} we show an example for the \texttt{network} block of an emulator trained on primary CMB $C_\ell^{TT}$ data for $2 \le \ell \le 10000$. We discuss the choices made in this block in more detail in \cref{subsec:network-specs-training}. 

\begin{figure}
    \begin{lstlisting}[language=yaml]
networks:
  - quantity: "Cl/tt"
    inputs: [ombh2, omch2, logA, ns, H0, tau]
    type: NN
    log: True
    modes:
      label: l
      range: [2,10000]
    n_traits:
      n_hidden: [512, 512, 512, 512]
    training:
      validation_split: 0.1
      learning_rates: [1.e-2, 1.e-3, 1.e-4, 1.e-5, 1.e-6, 1.e-7]
      batch_sizes: [1000, 2000, 5000, 10000, 20000, 50000]
      gradient_accumulation_steps: 1
      patience_values: 100
      max_epochs: 1000

    \end{lstlisting}
    \caption{Code snippet for network block. We setup a network that emulates $\log_{10}(C_\ell^{TT})(\vec{\theta})$ with our six input parameters $\vec{\theta} = \left\{ \Omega_b h^2, \Omega_c h^2, \log(10^{10} A_s), n_s, h, \tau \right\}$ and $\ell$ between 2 and 10000. The network is a fully connected dense neural network with 4 hidden layers of 512 neurons each. Our training block defines the fraction of example spectra used for validation estimation, the learning rates of each learning step, the batch size over which we average, any gradient accumulation steps, patience values, and maximum number of training epochs.}
    \label{snip:networks}
\end{figure}

\par Once the packaging file has been set up with the sections specified above, it becomes easy to generate training data for networks by calling:

\begin{lstlisting}[language=bash]
python -m cosmopower generate <yamlfile>
\end{lstlisting}

\par In addition, the \texttt{-{}-resume} flag can be used to increase more samples for an already existing set of data points, if it is found afterwards that the training set size is not large enough for training to result in good recovery of spectra from the emulator. When resuming the generation of samples, any pre-existing LHC will be used (if compatible with the given prescription) and any pre-existing samples are not re-generated. This can be used for continuing a run that was cancelled or stopped before, adding new quantities that were not computed earlier, or increasing the number of samples beyond the LHC that was generated beforehand.

\par We store the generated training data in hdf5 files, which are optimised for large, table-like datasets, and allow for both fast read-write access and good data compression. We also include the option to automatically split the data into multiple files, to prevent memory issues from opening a too large a single file at once. For our $\Lambda$CDM emulators, this means that we generate about 4 GB worth of training spectra per emulator, split across ten files.

\subsection{Network Specification and Training} \label{subsec:network-specs-training}

\par The \texttt{networks} block contains information on which emulator is to be trained, and how the network is designed; it contains:

\begin{enumerate}[leftmargin=*]
    \item The type of emulator, either \texttt{NN} for a neural network emulating the spectra directly, or \texttt{PCAplusNN} for a NN emulating the PCA of the quantity;
    \item The list of inputs used for the network, these can be different from the inputs to the Boltzmann code, and hence may need to be specified again;
    \item Whether the network should be trained on $\mathrm{log}$-spectra;
    \item The range of modes (sampling points) over which the output spectrum is computed, and a text label for them (i.e. $\ell$s for $C_\ell$ spectra, $k$ for $P(k)$ spectra, and redshifts $z$ for background quantities);
    \item The specification for traits of the Neural Network emulator. For a dense neural network, the traits should contain the number of nodes per hidden layer. For a network that employs a PCA, the number of retained PCs must be given.
    \item The specification for the steps taken when training the emulator (see below for details).
\end{enumerate}

\par After the training data has been generated, training a network is done via a similar command:

\begin{lstlisting}[language=bash]
python -m cosmopower train <yamlfile>
\end{lstlisting}

\par Training depends on a variety of parameters, which are set in the \texttt{training} block of the networks prescription. These parameters (explained below) are:

\begin{enumerate}[leftmargin=*]
    \item The learning rate, which controls the size of steps taken at each learning epoch;
    \item A batch size, which controls the size of a batch over which a learning step is averaged;
    \item The validation split, which controls how many spectra are kept aside of validation calculation;
    \item The number of steps used for gradient accumulation;
    \item A patience value, which controls how long a network allows itself to be ``stuck'' at a loss value before continuing to the next learning iteration;
    \item The maximum number of epochs in each learning iteration.
\end{enumerate}

\par Each of these values can be set to either a single number or a list of length $N_L$, which indicates the number of \emph{learning iterations} used. If a value is set to a single number, it is kept fixed over the course of each learning step, otherwise \cosmopower will iterate over the values in the list when training. If multiple values are to be iterated over, these lists need to be of the same length.

\par \cosmopower will train a network by iterating over these \emph{learning iterations}, each of which consists of a number of \emph{epochs} set by the \texttt{max\_epoch} value. A fraction of samples equal to the \texttt{validation\_split} is set aside each learning iteration, and the remainder is used as the training set. The training set is then grouped into \emph{batches} determined by the \texttt{batch\_size} value. Every epoch, each batch is passed through the emulator, and the trainable hyperparameters of the emulator are updated to reduce the loss function of the network. If a number of \texttt{gradient\_accumulation\_steps} $g > 1$ is given, then $g$ consecutive steps are used to compute the total derivative of the loss function with respect to the hyperparameters as well, which can give a better learning rate, especially when using a GPU for increased computation of these derivatives. \cosmopower uses the \emph{Adam} optimiser to determine how to tweak the hyperparameters, and the learning step size is multiplied by the \texttt{learning\_rate} of this iteration. After going through a full epoch, the validation set is passed through the emulator and its loss is computed. If the validation loss has improved throughout this iteration, then the new hyperparameters are kept. If the \texttt{max\_epoch} value is reached, or if the validation loss has not improved over \texttt{patience\_values} epochs in a row, then the emulator will go to the next learning iteration. 

\par Because of the large amount of freedom in choosing these values, it can be hard to determine what settings are optimal for a good training pass. In addition, the impact of certain decisions can wildly vary from either minimal to substantial. As a result, we cannot provide clear guidance on what settings to use but there are a few rules of thumb that can be used when determining the training settings which we recommend:

\begin{itemize}[leftmargin=*]
    \item The validation split should be about 10-20\%;
    \item Each iteration, the learning rate should go down and the batch size should go up;
    \item If a learning iteration reaches the maximum number of epochs instead of a patience value, that means it could have learned for longer, and it hasn't fully optimised yet - try to increase the batch size or learning rate for this iteration or an earlier one.
\end{itemize}

\par \cosmopower keeps track of the validation loss for every epoch, and saves this to a plain text file for post-training analysis and diagnosis of training issues.

\subsection{Assessing Accuracy} \label{subsec:accuracy-plotting}

\par The validation loss for the emulators is only one quantity to evaluate the accuracy, but it is important to explicitly evaluate the accuracy of the output emulator quantities. We include functionality to generate accuracy plots, that show the average difference between the emulated quantity and the original quantity as computed by the Einstein-Boltzmann code, relative to (`in units of') an observable error.

\par For a trained emulator, one can evaluate the accuracy of the emulator by invoking the command:

\begin{lstlisting}[language=bash]
python -m cosmopower show-validation <yamlfile>
\end{lstlisting}

\par This command will pass a fraction of all original samples through the each trained emulator and plot the emulator error. The accuracy of the emulated observables can be defined as either the fractional difference to the true value, or relative to some observational error, as defined in e.g.~\cref{eq:cmb-noise-curve}. There are options to use either the public Simons Observatory noise curves noise curves, presented in~\cite{SO_Noise_Model_2019}, or a cosmic variance-limited uncertainty.

\section{Wrapper Description} \label{sec:wrapper-description}

\par As an additional component for our \cosmopower extension, we provide wrapper functionality that interfaces the basic \cosmopower functionality with the inference software packages \cosmosis and \cobaya. Because most of the emulator specification will be present in the packaging prescription file, interfacing these emulators with the sampling software is as simple as pointing the wrapper to a packaging file. The remaining interfacing is then provided for with these wrappers. We will show here how to interface the emulators with \cosmosis and \cobaya, and show that these wrappers, with the emulators we have described in the previous section, can recover parameter constraints equivalent to those recovered with the original Einstein-Boltzmann code.

\subsection{\cosmosis} \label{subsec:wrapper-cosmosis}

\par The wrapper for using \cosmopower in \cosmosis inference pipelines involves specifying the \cosmopower module in the usual way in the \texttt{ini} file:

\begin{lstlisting}[language=yaml]
[cosmopower]
file = path/to/interface/cosmopower_interface.py
package_file = /path/to/packaging/package_prescription.yaml
extra_renames = {'cosmosis_parameter_name' :
                        'network_parameter_name'}
\end{lstlisting}

The options available and their default values for the module are specified in its associated \texttt{module.yaml}. In particular we note that care should be taken with parameter naming conventions, with any necessary translations specified using the \texttt{extra\_renames} keyword. The \cosmosis wrapper allows for the use of \cosmopower to compute CMB and matter power spectra, and the background evolution and derived quantities also described in \cref{subsec:emulated-quantities}. If desired, it is also possible to use \cosmopower only to perform the computation of spectra from the perturbations, and the native Einstein-Boltzmann code for the (relatively) faster background calculations (e.g. by only requesting the CMB from \cosmopower and including a \camb module with \texttt{mode = background}). Here we note that caution should be taken to not generate inconsistent results through inconsistent choices of \camb parameters when running in this mode.

\subsection{\cobaya} \label{subsec:wrapper-cobaya}

\par When \cosmopower is installed, the wrapper for using it in \cobaya can be used by simply adding the \texttt{cosmopower} block to the \cobaya configuration file. This is similar to how one normally adds \camb or \class as their Einstein-Boltzmann code. Due to the new interface using the packaging prescription, the \cosmopower wrapper requires minimal settings, and a full block can look as simple as: 

\begin{lstlisting}[language=yaml]
cosmopower:
  root_dir: /path/to/packaging
  package_file: package_prescription.yaml
\end{lstlisting}

\par Here, the (optional) \texttt{root\_dir} keyword points the wrapper to the root directory where the packaging file is saved, and the \texttt{package\_file} option points to the packaging prescription file that you want to load in. From this point, the wrapper parses the packaging prescription, interfaces with \cobaya, loads in the emulators that are required to compute all desired quantities, and provides the likelihoods with the computed quantities during the chain sampling.

\subsection{Fall through to native Einstein-Boltzmann code}

In order to increase the robustness of the use of \cosmopower emulators, we also include a feature which allows a given evaluation to `fall through' to the native Einstein-Boltzmann code, in a limited and configurable set of circumstances. By specifying the \texttt{fall\_through = True} option in the wrapper being used, \cosmopower will check that a python module corresponding to the \texttt{emulated\_code} and \texttt{version} can be imported. If so, then if a set of parameters is requested by the sampler which is outside of the trained range of the emulator specified in the \texttt{parameters} block (e.g. if the prior being used is wider than the training range) then \cosmopower will give a warning, but also calculate the requested quantities using the native Einstein-Boltzmann code. Whilst this may be desirable in a limited set of circumstances, care should be taken that the expected computational cost does not overwhelm that of augmenting the training set with a broader range of parameters and re-training the emulator.

\section{Comparison of Recovered Cosmology} \label{sec:recover-cosmology}

\par We now demonstrate that we can use our emulators in parameter inference analysis, generating posterior samples using Monte Carlo chains with each of the \cobaya and \cosmosis wrappers above using the same packaged network. In order to utilise all of the output quantities we do this for a set of observables: primary CMB, CMB lensing, galaxy weak lensing, and galaxy clustering. Note that this allows for quick and easy cross-validation of the results from using different Einstein-Boltzmann codes between different inference packages (e.g., \class in \cosmosis and \camb in \cobaya). This is particularly important because leading cosmology collaborations adopt different combinations of these codes while releasing results which we want to compare and combine.

\subsection{Simulated data vectors}

\par For full validation, it is important to check that not only the emulators recover the cosmological observables to high accuracy, but also that there is no inherent bias when using our emulators for estimation of the final cosmological parameters. To do this, we can generate simulated data for the observables we emulate with a theoretical covariance matrix and perform a parameter inference analysis on them using the wrappers described above. 

\begin{table}
    \centering
    \begin{tabular}{c c}
        \toprule
        \textbf{Parameter} & \textbf{Fiducial Value} \\
        \midrule
        $\Omega_b h^2$ & $2.2383 \times 10^{-2}$ \\
        $\Omega_c h^2$ & $12.011 \times 10^{-2}$ \\
        $H_0$ & $67.32 \, \mathrm{km}/\mathrm{s}/\mathrm{Mpc}$ \\
        $n_s$ & $0.966$ \\
        $\log(10^{10} A_s)$ & $3.0448$ \\
        $\tau$ & $5.43 \times 10^{-2}$ \\
        \midrule
        $A_b$ & $3.13$ \\
        $\eta_b$ & $0.603$ \\
        $\log T_\mathrm{AGN}$ & $7.8$ \\
        \midrule
        $\Sigma m_\nu$ & $0.12 {\rm eV}$ \\
        \bottomrule
    \end{tabular}
    \caption{The fiducial parameters used for generating the smooth data vector. The first six parameters refer to the cosmology, while the middle three are the baryonic feedback parameters used in the non-linear model of \camb. The last parameter is specific for the extension model we tested, with a neutrino mass for the inverted hierarchy to ensure that we could recover a closed posterior for our $+\Sigma m_\nu$ emulators. The remaining accuracy settings are the same as in~\cref{snip:accuracy}.}
    \label{tab:fiducial-cosmology}
\end{table}
\subsubsection{Cosmic-variance-limited CMB data}
\par For our testing purposes, we generate a smooth data vector with cosmic-variance-limited noise (such that our conclusions apply to all current and future experiments). This data vector contains data from a fiducial cosmology (see~\cref{tab:fiducial-cosmology}) for the CMB power spectra $C_\ell^{TT}, C_\ell^{TE}$, and $C_\ell^{EE}$, as well as the lensing potential spectrum $C_\ell^{\phi\phi}$. For the CMB data vector, the cosmic-variance-limited noise model is similar to~\cref{eq:cmb-noise-curve}, with $N_\ell^{XX} = N_\ell^{XY} = 0$ for all combinations of $XX$ and $XY$. We constrain our analysis to the multipole range $2 \le \ell \le 6000$. To explore the parameter space we add a log-likelihood function as a simple Gaussian chi-square distribution:

\begin{equation}
    \log \mathcal{L} = - \frac{1}{2} \sum_\ell \left( \frac{C_\ell^\mathrm{pred} - C_\ell^\mathrm{data}}{\sigma_\ell} \right)^2 .
\end{equation}

\par Since the data vector is smooth, we expect to recover the exact input parameters with a final $\chi^2 = 0$.

\subsubsection{Stage-IV-like 3x2pt LSS data}
\label{subsec:lss_data}
We also simulate a Large Scale Structure dataset for demonstrating and validating the $P(k)$ emulators. This consists of 3x2pt data for cosmic shear, galaxy clustering and galaxy-galaxy lensing, as is typically constrained by experiments such as DES, HSC and KiDS+BOSS+2dFLens. Here we approximate the constraining power of a Stage-IV LSS survey (such as LSST or \emph{Euclid}), with a number of caveats. In order to be able to make use of existing theoretical modelling and likelihoods which are implemented in \emph{both} \cobaya and \cosmosis we use real space data rather than power spectra and set up the redshift and angular binning of the data to be the same as the DES-Y1 configuration, as described in \cite{DES:2017myr}. Likewise, we both simulate and model the data using the DES-Y1 model for Intrinsic Alignments, linear galaxy bias, shear and redshift calibration biases etc. For a covariance matrix we create a Gaussian covariance using the \texttt{save\_2pt} module of \cosmosis. We do not contend such a model will be accurate for describing real Stage-IV data; here we are seeking to understand if differences between the calculation of $P(k)$ with either \cosmopower or \camb can be detected when 3x2pt statistics are measured with Stage-IV precision. To that end we assume a sky fraction, redshift distribution, total galaxy number density and shape noise as appropriate for an LSST-Y10 3x2pt survey \citep[as specified in the LSST-SRDC by][]{LSSTDarkEnergyScience:2018jkl} when simulating and analysing the data. Full details of the configuration are given in \cref{sec:3x2-appendix}.

\subsection{Results}

\par
\begin{figure*}
    \includegraphics[width=0.5\textwidth]{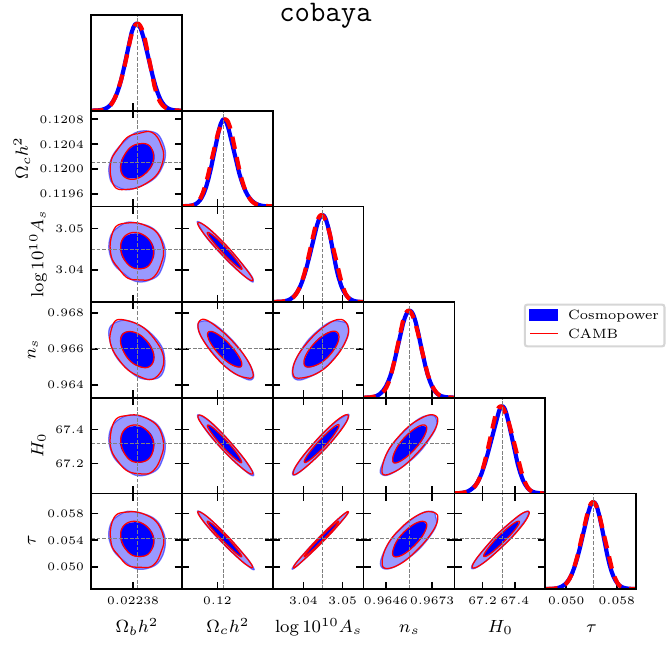}\includegraphics[width=0.5\textwidth]{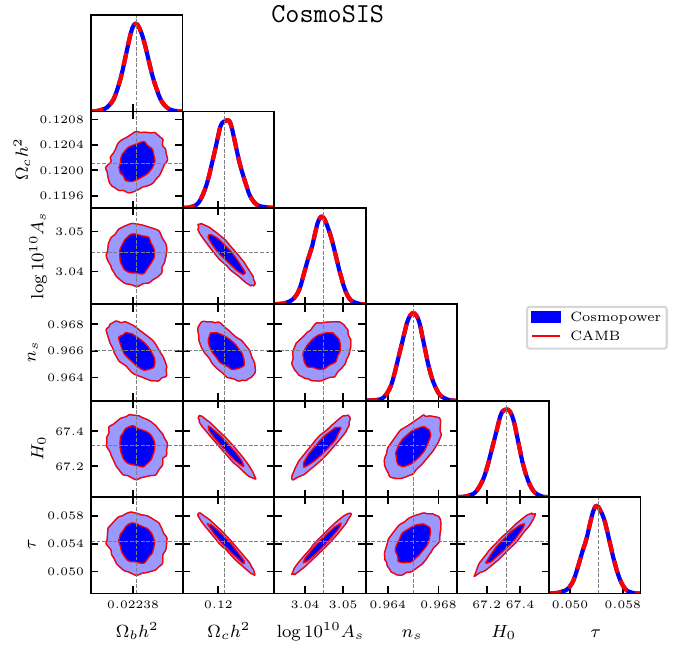}
    \caption{To illustrate that we can estimate posteriors in both \cobaya and \cosmosis we show the same 68\% and 95\% confidence levels for \lcdm parameters from CMB cosmic-variance-limited power spectra, obtained from a full MCMC run done either with the \cobaya wrapper for \cosmopower (\textbf{blue}) or with the  \camb (\textbf{red}) on the \emph{left} for \cobaya and \emph{right} for \cosmosis. This Figure also demonstrates the correct recovery of the cosmological likelihood in each case (note that for \cobaya two separate sets of posterior samples are taken, whilst for \cosmosis we re-evaluate the likelihood at the same posterior samples, resulting in visually identical contours).}
    \label{fig:cobaya-wrapper-comparison}
\end{figure*}

 \cref{fig:cobaya-wrapper-comparison} shows the recovered contours of \cobaya\unskip+\cosmopower and \cosmosis\unskip+\cosmopower versus the \cobaya\unskip+\camb and \cosmosis\unskip+\camb posteriors from a CMB cosmic-variance-limited dataset. We show that we can reproduce the \camb best-fit cosmology and posterior distribution to $< 0.1 \sigma$ of the cosmic variance limit error bars in both inference codes. \Cref{{fig:cobaya-mnu-wrapper-comparison}} shows the same result within \cobaya for the $+\sum m_{\nu}$ emulator as an example for an extended model.

\par The main advantage from running \cosmopower is the speed increase over \camb. For a simple $\Lambda$CDM model and the cosmic-variance-limited CMB data, we found that a \camb chain took $\sim$ 10 hours, while for \cosmopower it takes only $\sim$ 20 minutes to run to convergence. Most of this speed-up comes from the fact that at this level of accuracy, an evaluation of a \camb power spectrum takes $\sim 20 s$ to compute, while the same computation takes \cosmopower $\sim 0.1 s$, at which point computing any non-trivial likelihood function becomes the limiting factor. When going to beyond-$\Lambda$CDM models, the time it takes to run a \camb chain will go up due to the increased complexity or accuracy requirements from the computations. For \cosmopower however, the pre-trained emulators do not require more complicated computation when running these chains, and as such the time it takes a \cosmopower chain to converge will only increase slightly due to the larger parameter space that needs to be explored.

\begin{figure}
    \includegraphics[width=0.5\textwidth]{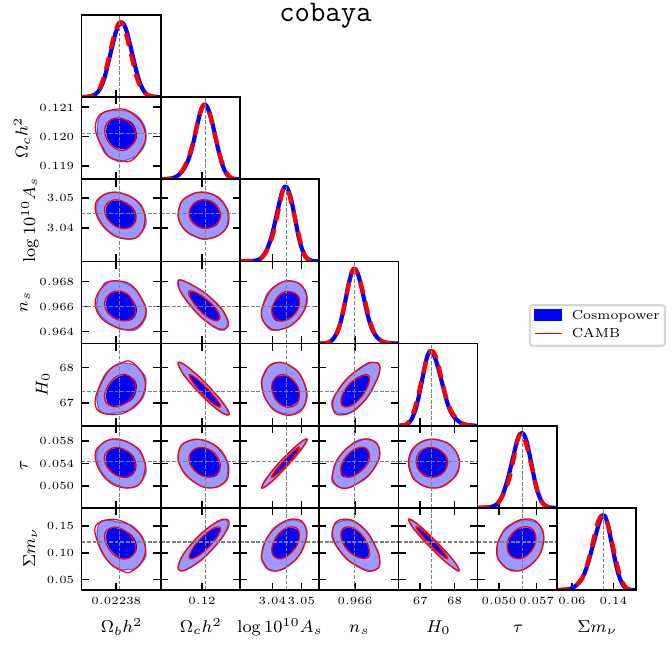}
    \caption{Similar to~\cref{fig:cobaya-wrapper-comparison} but for \lcdm$+\Sigma m_\nu$: \emph{Left:} 68\% and 95\% confidence levels for \lcdm parameters from CMB cosmic-variance-limited power spectra, obtained from a full MCMC run done either with the \cobaya wrapper for \cosmopower (\textbf{blue}) or with the existing \camb wrapper for \cobaya (\textbf{red}). The dotted lines show the fiducial value of the input data vector, and both posterior distributions recovered this fiducial value within $< 0.1 \sigma$. The \cosmopower sampler converged within $\sim$ 100 minutes, while the \camb sampler converged after $\sim$ 28 hours.}
    \label{fig:cobaya-mnu-wrapper-comparison}
\end{figure}

\begin{figure}
    \includegraphics[width=0.5\textwidth]{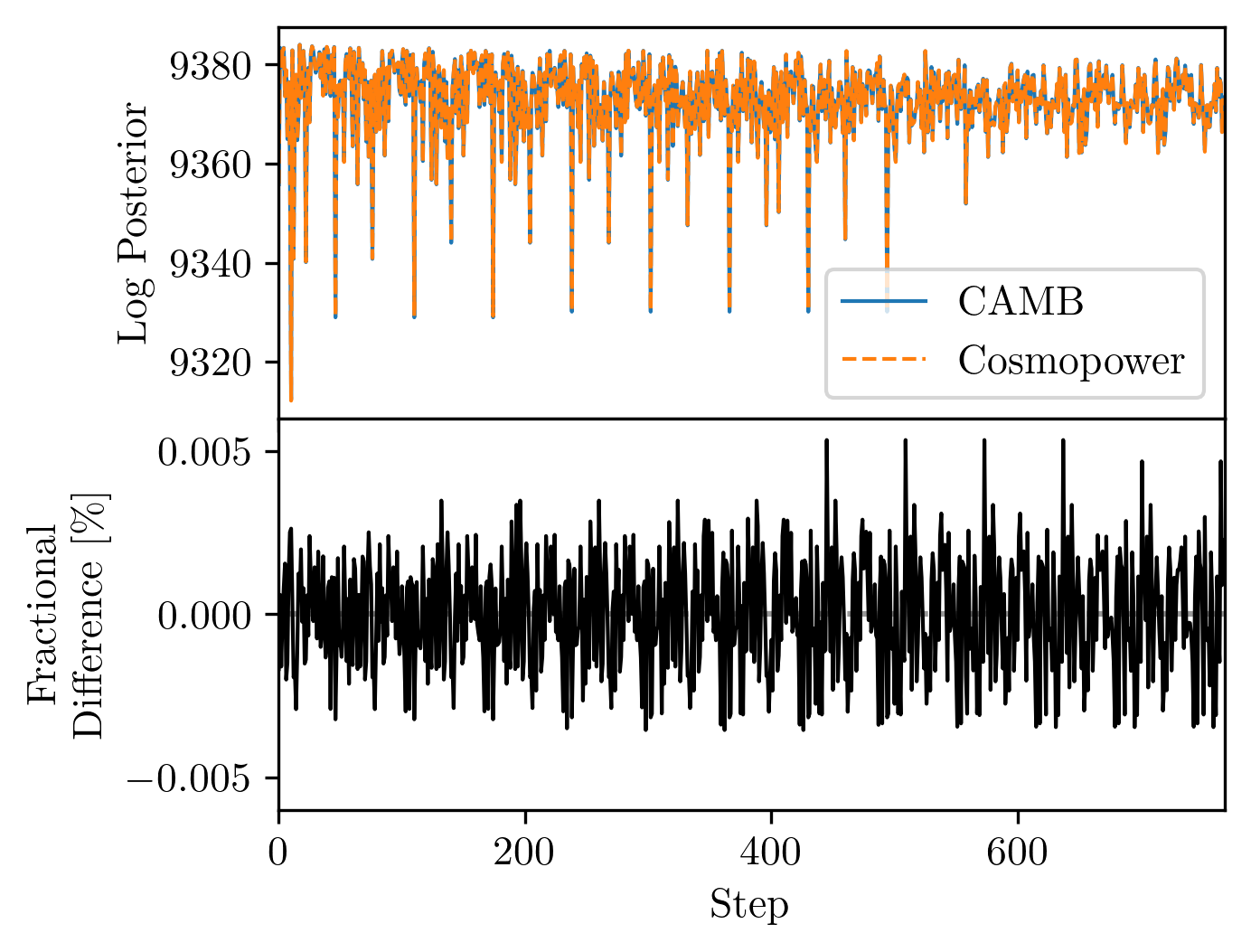}
    \caption{Log Posterior differences for the Stage-IV-like 3x2pt LSS data set described in \cref{subsec:lss_data} between estimations made using the original \camb Boltzmann code and the \cosmopower emulator.}
    \label{fig:3x2trace}
\end{figure}

Similarly for the Stage-IV-like LSS data we find times for each individual likelihood evaluation with the \cosmopower \cosmosis module to be $\sim0.5\,$seconds, compared to $\sim42\,$seconds for the \camb \cosmosis module. In this case the need for Limber integration dominates the likelihood evaluation time for \cosmosis ($\sim2\,$seconds) when the Boltzmann emulator is used. Rather than expending significant computational expense on a fully converged \camb chain, in \cref{fig:3x2trace} we show the log Posterior values calculated in a short chain using both \camb and \cosmopower within \cosmosis for the LSS data set described in \cref{subsec:lss_data}. As can be seen the relative differences in log Posterior between the numerical code and the emulator are less than $0.005\%$, representing an indistinguishable difference in estimates of posterior credible intervals and summary statistics. See \cref{fig:mock3x2} for full estimated posteriors showing the parameter constraining power of this data set.

\section{Conclusions}\label{sec:conclusion}
\par We have presented a coherent framework for specifying, creating, packaging and utilising emulators of cosmological Einstein-Boltzmann codes, building on the \cosmopower package. These emulators can speed up by orders of magnitude the estimation of posteriors on cosmological and nuisance parameters from experimental data and hence enable investigation of models which extend the fiducial \lcdm cosmology and the checking of the robustness of any conclusions made to a plethora of modelling choices. By creating a specification for packaging and distributing such emulators and providing wrappers for their use in popular inference packages we hope to improve efficiency and reproducibility in cosmological studies, by allowing appropriate emulators to be widely used by many different studies once they have been trained. This kind of reproducibility across platforms will also assist in combining different data sets to improve statistical constraining power and investigate more models in more detail.

We have used the framework to produce a suite of emulators of quantities calculated by \camb v1.5.0: CMB primary angular power spectra $C_\ell^{TT}, C_\ell^{TE}, C_\ell^{EE}, C_\ell^{BB}$; CMB lensing power spectra $C_\ell^{\phi\phi}$; linear and non-linear matter power $P(k)_\mathrm{lin}$, $P(k)_\mathrm{NL}$ and a variety of background and derived quantities. We have demonstrated the accuracy of the emulators at both the spectrum level and the parameter-recovery level to accuracy appropriate for Stage-IV data (and beyond to the cosmic variance limit for the CMB spectra).

In principle, this standardisation of emulator packaging extends in scope beyond Einstein-Boltzmann codes to other numerically-intensive codes amenable to emulation, such as Interstellar Medium models \citep[e.g.][]{2023A&A...678A.198P}, supernova radiative transfer \citep[e.g.][]{2021ApJ...910L..23K}, early-Universe re-ionisation models \citep[e.g.][]{Schmit:2017pho} and others.

The framework described here will form a new release of the \cosmopower code, with the website \url{https://alessiospuriomancini.github.io/cosmopower/} providing full API documentation and extensive demo scripts and tutorial notebooks.

\section*{Acknowledgements}
HTJ, IH and EC acknowledge support from the European Research Council (ERC) under the European Union's Horizon 2020 research and innovation programme (Grant agreement No. 849169). ASM acknowledges support from the MSSL
STFC Consolidated Grant ST/W001136/1. We acknowledge the support of the Supercomputing Wales project, which is part-funded by the European Regional Development Fund (ERDF) via Welsh Government. We thank Antony Lewis for input on precision parameters and support with \cobaya; Jens Chluba and Yacine Ali-Ha\"{i}moud for discussions on recombination codes; and Joe Zuntz for discussions on \cosmosis. In addition to the references in the main text we thank the authors and maintainers of public software codes including \texttt{NumPy}~\citep{harris_array_2020}, \texttt{SciPy}~\citep{virtanen_scipy_2020}, \texttt{matplotlib}~\citep{hunter_matplotlib_2007}, \texttt{TensorFlow}~\citep{tensorflow2015-whitepaper}, and \texttt{GetDist}~\citep{Lewis:2019xzd}.

\subsubsection*{Author contributions}
\par We list here the roles and contributions of the authors according to the Contributor Roles Taxonomy (CRediT)\footnote{\url{https://credit.niso.org/}}. \newline
\textbf{Hidde T. Jense}: Conceptualization (equal), Investigation (equal), Methodology (equal), Software (lead), Validation (equal), Visualization (lead), Writing - original draft (equal). \textbf{Ian Harrison}: Conceptualization (equal), Investigation (equal), Methodology (equal), Software (supporting), Supervision (supporting), Validation (equal), Visualization (supporting), Writing - original draft (equal). \textbf{Erminia Calabrese}: Conceptualization (equal), Methodology (supporting), Supervision (lead), Visualization (supporting), Writing - original draft (equal). \textbf{Alessio Spurio Mancini}: Conceptualization (equal), Methodology (equal), Writing - original draft (supporting). \textbf{Boris Bolliet}: Conceptualization (equal), Writing - original draft (supporting). \textbf{Jo Dunkley}: Writing - original draft (supporting). \textbf{J. Colin Hill}: Conceptualization (supporting), Writing - original draft (supporting).

\bibliographystyle{mnras_2author}
\bibliography{citations.bib}

\appendix

\FloatBarrier
\section{Full $\Lambda$CDM Emulator Prescription} \label{sec:lcdm-prescription}

\par Here we present the full yaml prescription for our $\Lambda$CDM emulators.

\begin{lstlisting}[language=yaml]
network_name: jense_2023_camb_lcdm
path: jense_2023_camb_lcdm

# Details on the boltzmann code we emulate
emulated_code:
  name: camb
  version: "1.5.0"
  inputs: [ ombh2, omch2, As, ns, H0, tau ]
  extra_args:
    lens_potential_accuracy: 8
    kmax: 10.0
    k_per_logint: 130
    lens_margin: 2050
    AccuracyBoost: 1.0
    lAccuracyBoost: 1.2
    lSampleBoost: 1.0
    DoLateRadTruncation: false
    min_l_logl_sampling: 6000
    recombination_model: CosmoRec

# Details on the parameters we sample and derive.
samples:
  Ntraining: 100000

  parameters:
    # Our latin hypercube
    ombh2: [0.015,0.030]
    omch2: [0.09,0.15]
    logA: [2.5,3.5]
    tau: [0.02, 0.20]
    ns: [0.85, 1.05]
    h: [0.4,1.0]
    # Parameters derived directly from our LHC
    H0: "lambda h: h * 100.0"
    As: "lambda logA: 1.e-10 * np.exp(logA)"
    # Parameters computed by our boltzmann code
    thetastar:
    sigma8:
    YHe:
    zrei:
    taurend:
    zstar:
    rstar:
    zdrag:
    rdrag:
    N_eff:

# Details on each of the emulators we want to create.
networks:
  - quantity: "derived"
    type: NN
    n_traits:
      n_hidden: [ 512, 512, 512, 512 ]
    training:
      validation_split: 0.1
      learning_rates: [ 1.e-2, 1.e-3, 1.e-4, 1.e-5, 1.e-6, 1.e-7 ]
      batch_sizes: [ 1000, 2000, 5000, 10000, 20000, 50000 ]
      gradient_accumulation_steps: [ 1, 1, 1, 1, 1, 1 ]
      patience_values: [ 100, 100, 100, 100, 100, 100 ]
      max_epochs: [ 1000, 1000, 1000, 1000, 1000, 1000 ]

  - quantity: "Cl/tt"
    type: NN
    log: True
    modes:
      label: l
      range: [2,10000]
    n_traits:
      n_hidden: [ 512, 512, 512, 512 ]
    training:
      validation_split: 0.1
      learning_rates: [ 1.e-2, 1.e-3, 1.e-4, 1.e-5, 1.e-6, 1.e-7 ]
      batch_sizes: [ 1000, 2000, 5000, 10000, 20000, 50000 ]
      gradient_accumulation_steps: [ 1, 1, 1, 1, 1, 1 ]
      patience_values: [ 100, 100, 100, 100, 100, 100 ]
      max_epochs: [ 1000, 1000, 1000, 1000, 1000, 1000 ]

  - quantity: "Cl/te"
    type: PCAplusNN
    modes:
      label: l
      range: [2,10000]
    p_traits:
      n_pcas: 512
      n_batches: 10
    n_traits:
      n_hidden: [ 512, 512, 512, 512 ]
    training:
      validation_split: 0.1
      learning_rates: [ 1.e-2, 1.e-3, 1.e-4, 1.e-5, 1.e-6, 1.e-7 ]
      batch_sizes: [ 1000, 2000, 5000, 10000, 20000, 50000 ]
      gradient_accumulation_steps: [ 1, 1, 1, 1, 1, 1 ]
      patience_values: [ 100, 100, 100, 100, 100, 100 ]
      max_epochs: [ 1000, 1000, 1000, 1000, 1000, 1000 ]

  - quantity: "Cl/ee"
    type: NN
    log: True
    modes:
      label: l
      range: [2,10000]
    n_traits:
      n_hidden: [ 512, 512, 512, 512 ]
    training:
      validation_split: 0.1
      learning_rates: [ 1.e-2, 1.e-3, 1.e-4, 1.e-5, 1.e-6, 1.e-7 ]
      batch_sizes: [ 1000, 2000, 5000, 10000, 20000, 50000 ]
      gradient_accumulation_steps: [ 1, 1, 1, 1, 1, 1 ]
      patience_values: [ 100, 100, 100, 100, 100, 100 ]
      max_epochs: [ 1000, 1000, 1000, 1000, 1000, 1000 ]
  
  - quantity: "Cl/bb"
    type: NN
    log: True
    modes:
      label: l
      range: [2,10000]
    n_traits:
      n_hidden: [ 512, 512, 512, 512 ]
    training:
      validation_split: 0.1
      learning_rates: [ 1.e-2, 1.e-3, 1.e-4, 1.e-5, 1.e-6, 1.e-7 ]
      batch_sizes: [ 1000, 2000, 5000, 10000, 20000, 50000 ]
      gradient_accumulation_steps: [ 1, 1, 1, 1, 1, 1 ]
      patience_values: [ 100, 100, 100, 100, 100, 100 ]
      max_epochs: [ 1000, 1000, 1000, 1000, 1000, 1000 ]
  
  - quantity: "Cl/pp"
    inputs: [ ombh2, omch2, logA, ns, h ]
    type: PCAplusNN
    log: True
    modes:
      label: l
      range: [2,10000]
    p_traits:
      n_pcas: 64
      n_batches: 10
    n_traits:
      n_hidden: [ 512, 512, 512, 512 ]
    training:
      validation_split: 0.1
      learning_rates: [ 1.e-2, 1.e-3, 1.e-4, 1.e-5, 1.e-6, 1.e-7 ]
      batch_sizes: [ 1000, 2000, 5000, 10000, 20000, 50000 ]
      gradient_accumulation_steps: [ 1, 1, 1, 1, 1, 1 ]
      patience_values: [ 100, 100, 100, 100, 100, 100 ]
      max_epochs: [ 1000, 1000, 1000, 1000, 1000, 1000 ]
\end{lstlisting}

\section{Dataset file structure} \label{sec:file-structure}

\par We opted to standardise the dataset file structure for \cosmopower, as a way to streamline the emulator building process for the end-user. At the \python-interface side, we included a \texttt{cosmpower.Dataset} class that wraps around the file structure easily and handles the file parsing in a safe manner.

\par The main file format we settled on is Hierarchical Data Format revision 5 (HDF5), which is a file format designed to handle large datasets of tabular nature, something that lends itself specially well for this issue. Via the \texttt{h5py} library in \python, HDF5 is also a relatively fast and memory-efficient read/write access, offering both good compression for hard drive storage and decompression rates for RAM access during runtime.

\par The training data needs to accurately match the $\vec{d}(\vec{\theta})$ mapping of our emulators well, while also being robust against potentially missing datapoints and multi-threaded reading access. We opted to split this mapping into two different files, a \texttt{parameters} file which contains the main LHC of the dataset and is only used for spectra generation, and a (set of) files for the computed observable quantities, which are named as \texttt{Cl\_tt.0.hdf5}, \texttt{Cl\_tt.1.hdf5}, etc. for e.g. $C_\ell^{TT}$. The quantity files are split into several files, to allow multi-threaded write access without having to worry about data races, and to prevent issues when opening data files which are larger than a device's available RAM.

\par The \texttt{parameters} file contains a header and a main body. The header contains an ordered list of strings for the $p$ parameters that are to be passed on to the Boltzmann code. The main body contains a $p \times N$ table of $N$ samples from the LHC. Because the LHC is relatively small in size and quick to generate, this file never needs to be written to in different threads and can be kept as one file. It is stored separately from the main dataset in case a spectra generation run is interrupted and needs to be resumed at a later stage, in which case it can be ensured that new spectra are sampled from the same LHC as before.

\par Each quantity file also contains a header and a main body. The header contains a list of $M$ modes for the quantity, the names of the parameters that are to be used for the emulator. In the main body, there is a $p \times N$ array of input parameters for each spectra, and a $M \times N$ array where each $M$-length spectrum is stored. In addition, there is a $N$-array of indices stored, the entries of which refer to the indices of the \texttt{parameters} file that each sample was computed from. Because quantity files are pre-allocated before they are filled, an index of -1 indicates that a spectrum has not been computed yet.

\section{Principal Component Analysis} \label{sec:pca-appendix}

\par The use of Principal Component Analysis (PCA) can be worthwhile in improving the accuracy of the emulator by compressing the full data into a smaller number of free components. While the reduction in freedom in the output is reduced and has therefore less capacity to accurately recover the original spectra, the reduced dimensionality of the output vector means that the emulator can more efficiently train on this reproduction.

\par The choice of whether or not to use PCA is not trivial, and there is no simple test that can conclusively show that the use of PCA compression is guaranteed to be beneficial before training an emulator. While for some cases, like $C_\ell^{TE}$, the use of a PCA is needed due to the zero-crossing of the observed quantity, it may not be obvious \emph{a priori} that the use of a PCA can improve it for other quantities as well.

\begin{figure}
    \centering
    \includegraphics[width=0.45\textwidth]{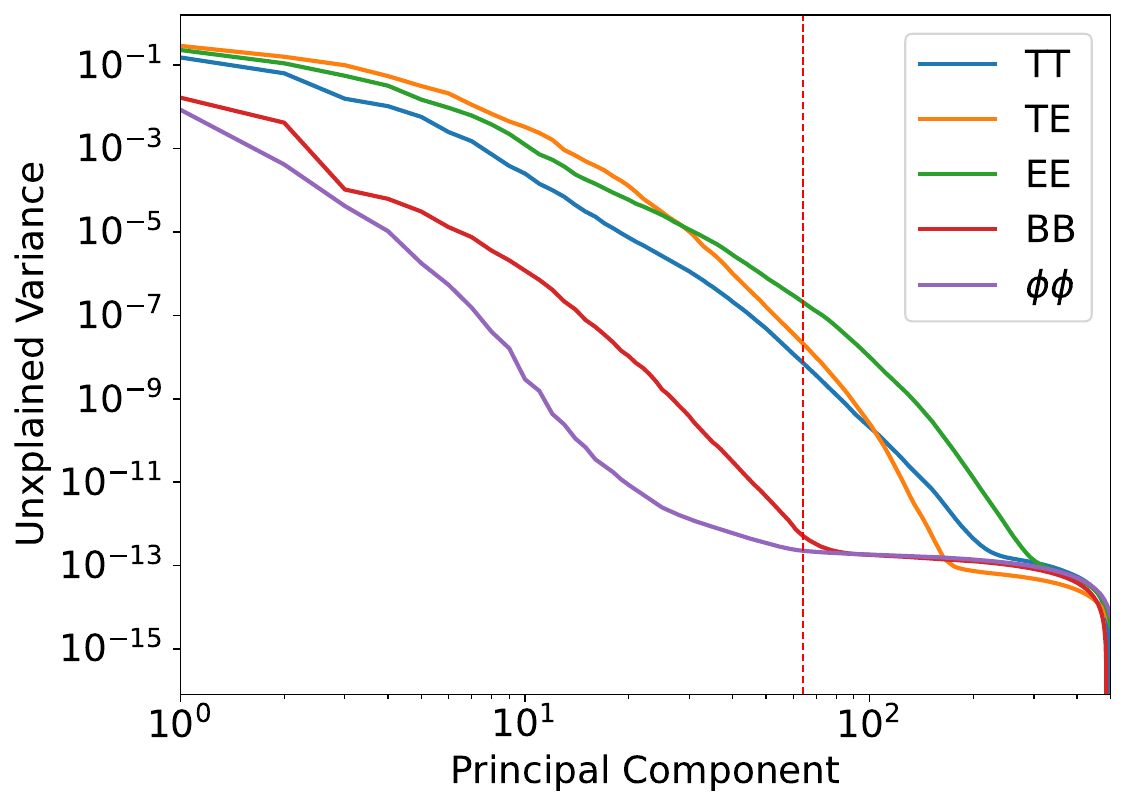}
    \caption{A scree plot, showing the \emph{unexplained variance} of a PCA compression for the various CMB quantities, as a function of the number of retained principal components. The ``scree'' of each line is the flat plateau of each line. We observed that for $C_\ell^{\phi\phi}$, this scree lies around 64 principal components (vertical red line), and hence a PCA compression of 64 retained components is effective for a $C_\ell^{\phi\phi}$ emulator. Conversely, however, observing the scree for $C_\ell^{BB}$ at around 100 principal components, we expected the same to see for this quantity. We attribute the lack of an improvement in emulation for this quantity to the presence of important features which shift in $\ell$-space for that quantity, which would not be retained by our implementation of PCA.}
    \label{fig:scree-plot}
\end{figure}

\par It was observed in~\cite{CosmoPower2021} that the $C_\ell^{\phi\phi}$ emulator improved in accuracy when employing PCA compression. We observed that this can be explained by making a \emph{scree} plot, which is a line plot of the eigenvalues of all retained PCA components. We show a scree plot of the training data for our $\Lambda$CDM emulators in~\cref{fig:scree-plot}. By observing where this line flattens out (the ``scree'' of the line), one can estimate the amount of components that need to be retained in the PCA. For the $C_\ell^{\phi\phi}$ spectra, we found that this scree appears around 60 components, which means around 64 components should be sufficient to accurately decompose the 10000 $\ell$ modes of the spectra without loss of information. Similarly, a scree plot showed that a few hundred components should be sufficient for $C_\ell^{TE}$. \\

\par However, a scree plot is not necessarily conclusive. We observed that the $C_\ell^{BB}$ are also dense enough that about 200 PCA components should be capable of accurately recovering them. Upon training such an emulator however, we found that direct emulation of $C_\ell^{BB}$ was more accurate than one that employed PCA compression. We think this is due to the fact that the BB spectra contain features which vary in $\ell$ under certain parameter variations, and hence cannot be properly accounted for in PCA compression. Since our regular emulators were shown to be more than accurate for physical analysis, we did not do an in-depth analysis of this discrepancy. Further investigation, or a different type of information compaction that does allow for horizontal shifts in $\ell$-space, can perhaps allow for more accurate emulators in the future.

\section{Specification of Stage-IV-like 3x2pt data} \label{sec:3x2-appendix}
For assessing the accuracy of our emulation of $P(k)$ at Stage-IV levels of precision on LSS data, we create a data set containing angular correlation functions for galaxy clustering $w(\theta)$, galaxy-galaxy lensing $\gamma_t(\theta)$, and cosmic shear $\xi_{\pm}(\theta)$. In addition to the fiducial cosmological model and parameters for \lcdm shown in \cref{tab:fiducial-cosmology}, we include linear galaxy bias parameters for the lens galaxies, a two-parameter NLA model for galaxy intrinsic alignments, one-parameter per tomographic bin central shift parameters for redshift distributions of the sources and lenses, and one parameter per tomographic bin for multiplicative shear bias calibration of the sources. Following the LSST-SRDC \citep{LSSTDarkEnergyScience:2018jkl} specification for LSST-Y10 we assume a redshift distribution for both sources and lenses given by $n(z) \propto z^2 \exp \left[ -(z / z_0)^\alpha \right]$ with $\alpha=0.783, z_0=0.176$ and convolve this with a Gaussian of width $\sigma_z = 0.05(1 + z)$. Sources are placed into four tomographic bins and lenses placed into five tomographic bins, all equally populated with the total number density of galaxies $n_{\rm gal} = 27 \, [\mathrm{arcmin}^{-2}]$ (note that this tomographic binning is not the one expected for the LSST analysis, but matches the DES-Y1 model). When modelling the covariance we assume a $\sigma_e = 0.26$ and a sky area of $14,300\,$deg$^2$. In \cref{fig:3x2trace} we show the \lcdm model constraints from this data set (using \cosmopower), alongside the offical Dark Energy Survey Y1 results from \cite{DES:2017myr} (which use the same model and likelihood pipeline) to give a sense of the relative power.
\begin{figure}
    \includegraphics[width=\columnwidth]{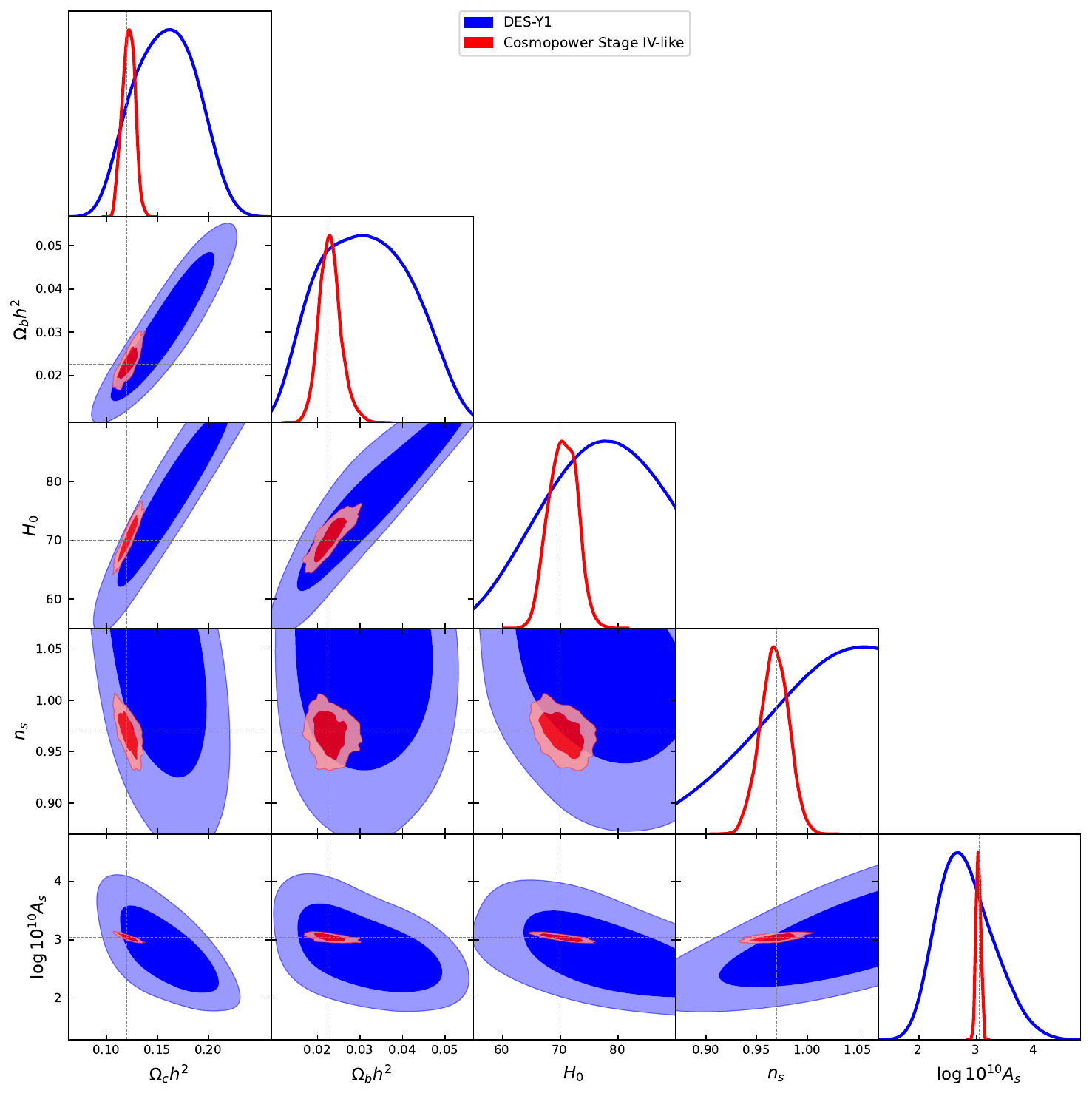}
    \caption{The \lcdm model constraining power of the Stage-IV-like 3x2pt Large Scale Structure data set used to benchmark the trained $P(k)$ emulator. For scale we show the official DES-Y1 \citep{DES:2017myr} chain, which use the exact same likelihood pipeline but with their real data.}
    \label{fig:mock3x2}
\end{figure}

\section{Accuracy plots for Extension model emulators} \label{sec:extension_accuracy}
Here we reproduce \cref{fig:lcdm-cmb-cl-validation} for the extended models we consider beyond \lcdm, with all models showing acceptable levels of accuracy as discussed in \cref{subsec:accuracy-validation}.

\begin{figure*}
    \centering
    \begin{subfigure}{0.4\textwidth}
        \includegraphics[width=\textwidth]{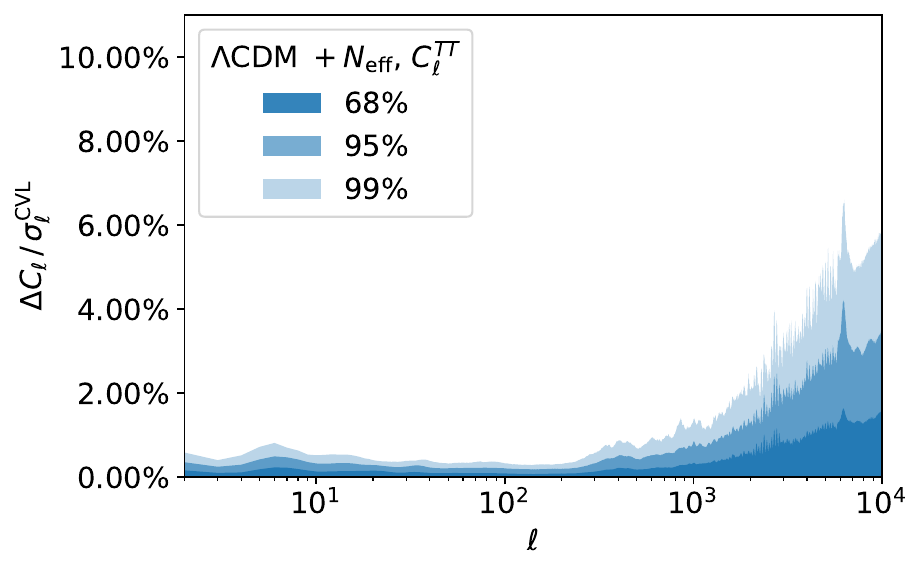}
        \vspace{-0.6cm}\caption{TT}
        \label{fig:neff-cvl-cl-tt}
    \end{subfigure}
    \begin{subfigure}{0.4\textwidth}
        \includegraphics[width=\textwidth]{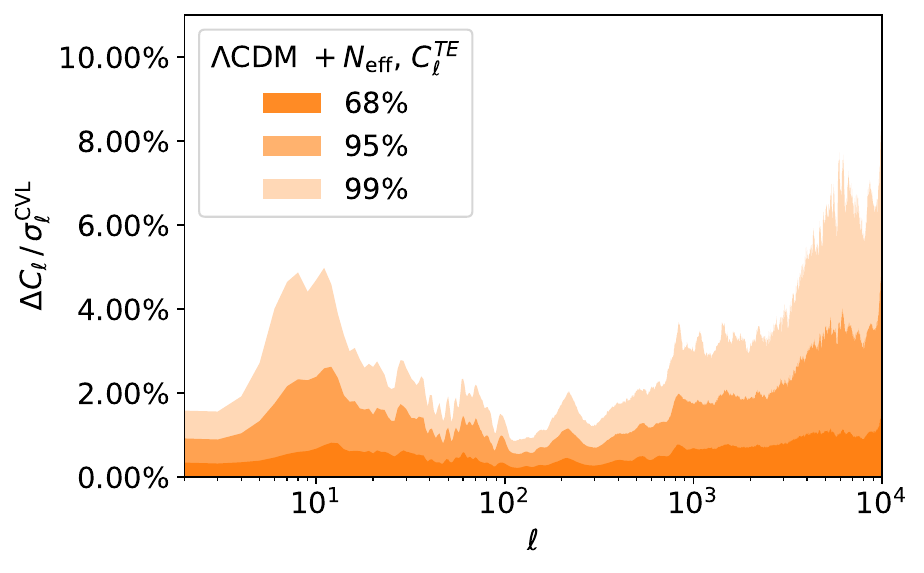}
        \vspace{-0.6cm}\caption{TE}
        \label{fig:neff-cvl-cl-te}
    \end{subfigure}
    \begin{subfigure}{0.4\textwidth}
        \includegraphics[width=\textwidth]{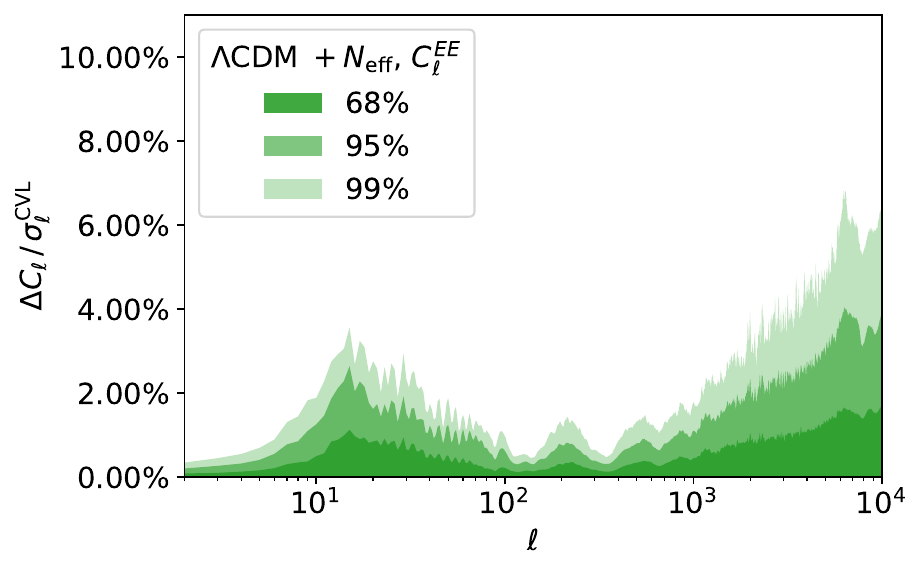}
        \vspace{-0.6cm}\caption{EE}
        \label{fig:neff-cvl-cl-ee}
    \end{subfigure}
    \begin{subfigure}{0.4\textwidth}
        \includegraphics[width=\textwidth]{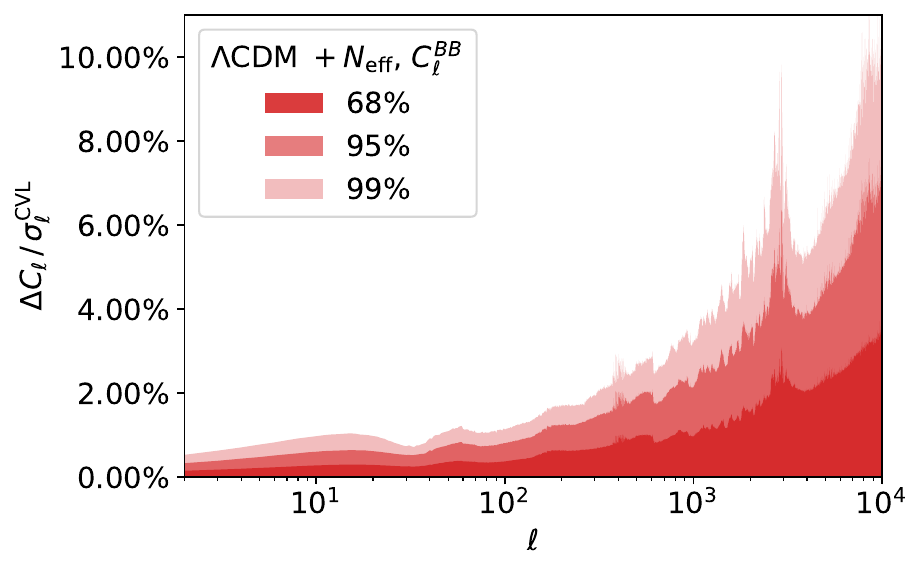}
        \vspace{-0.6cm}\caption{BB}
        \label{fig:neff-cvl-cl-bb}
    \end{subfigure}
    \begin{subfigure}{0.4\textwidth}
        \includegraphics[width=\textwidth]{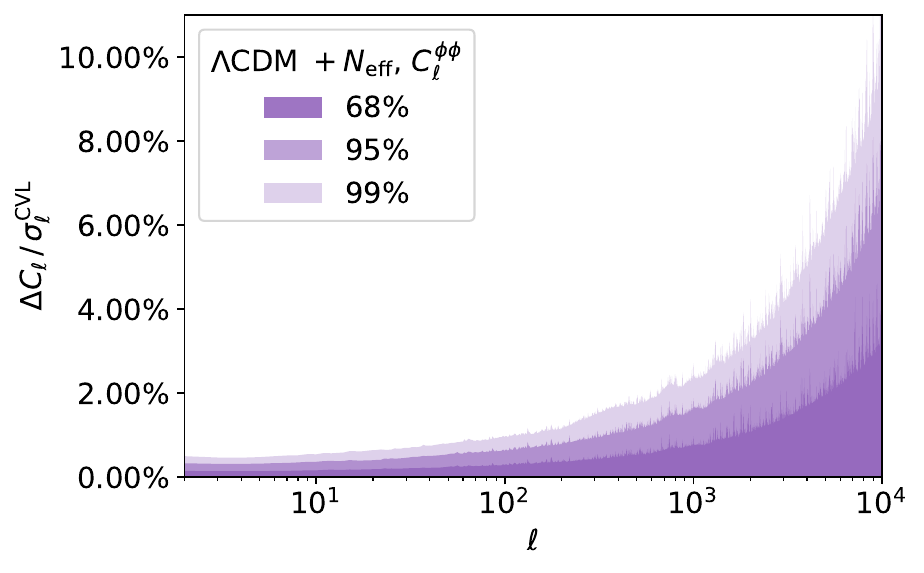}
        \vspace{-0.6cm}\caption{$\phi\phi$}
        \label{fig:neff-cvl-cl-pp}
    \end{subfigure}
    \begin{subfigure}{0.38\textwidth}
        \includegraphics[width=\textwidth]{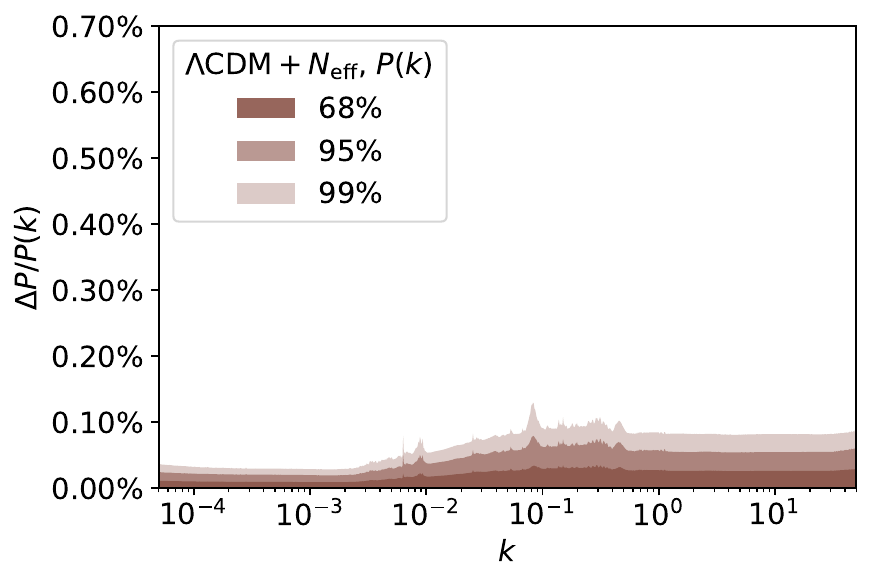}
        \vspace{-0.6cm}\caption{$P(k)$}
        \label{fig:neff-Pklin}
    \end{subfigure}
    \caption{Same as~\cref{fig:lcdm-cmb-cl-validation} but for $\Lambda$CDM${}+N_\mathrm{eff}$. }
    \label{fig:neff-cmb-cl-validation}
\end{figure*}

\begin{figure*}
    \centering
    \begin{subfigure}{0.4\textwidth}
        \includegraphics[width=\textwidth]{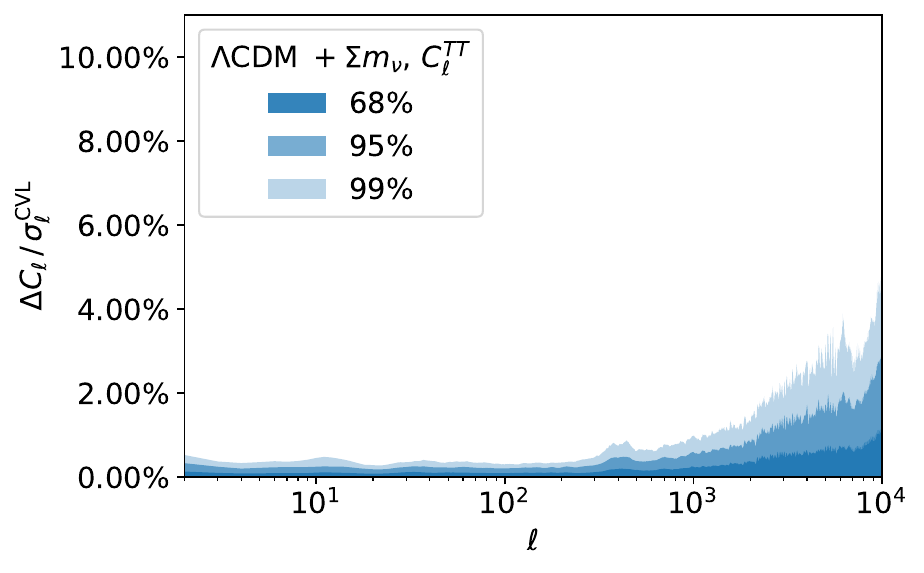}
        \vspace{-0.6cm}\caption{TT}
        \label{fig:mnu-cvl-cl-tt}
    \end{subfigure}
    \begin{subfigure}{0.4\textwidth}
        \includegraphics[width=\textwidth]{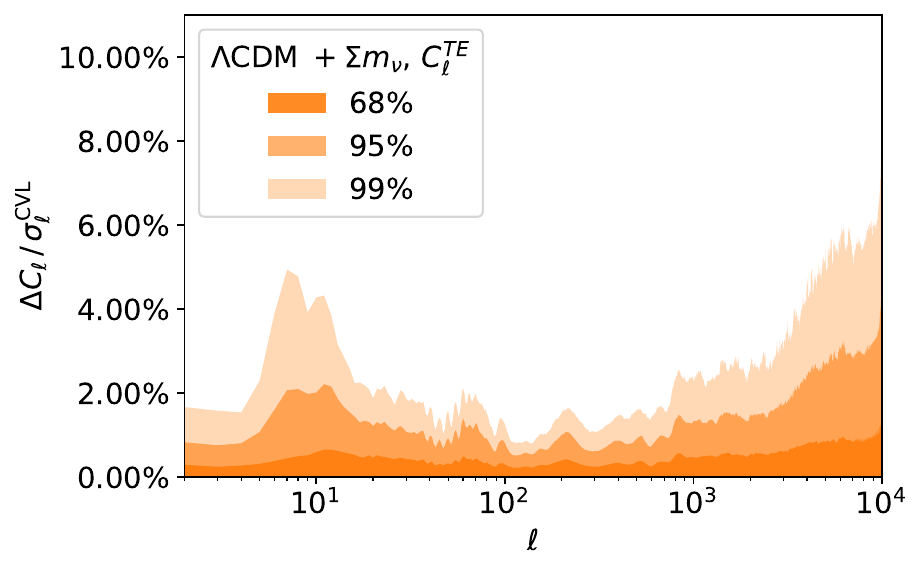}
        \vspace{-0.6cm}\caption{TE}
        \label{fig:mnu-cvl-cl-te}
    \end{subfigure}
    \begin{subfigure}{0.4\textwidth}
        \includegraphics[width=\textwidth]{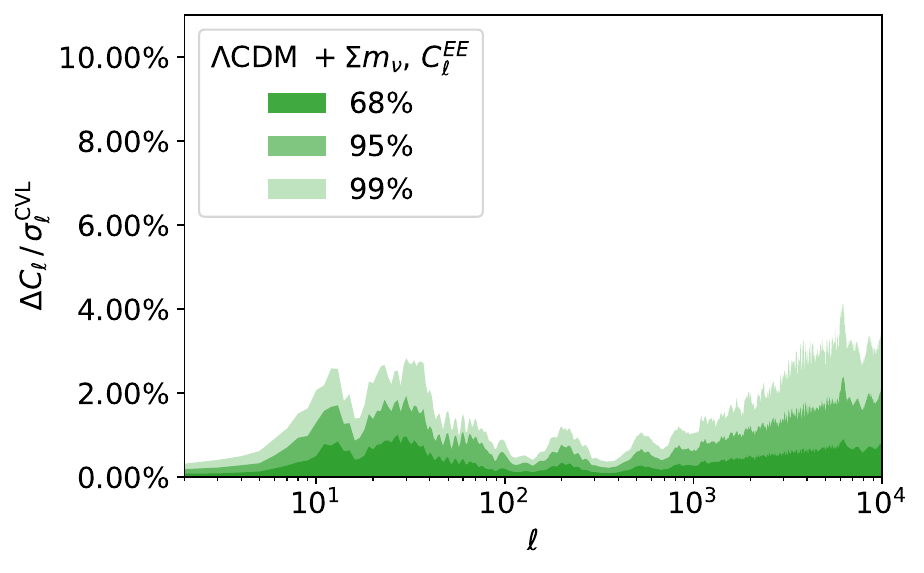}
        \vspace{-0.6cm}\caption{EE}
        \label{fig:mnu-cvl-cl-ee}
    \end{subfigure}
    \begin{subfigure}{0.4\textwidth}
        \includegraphics[width=\textwidth]{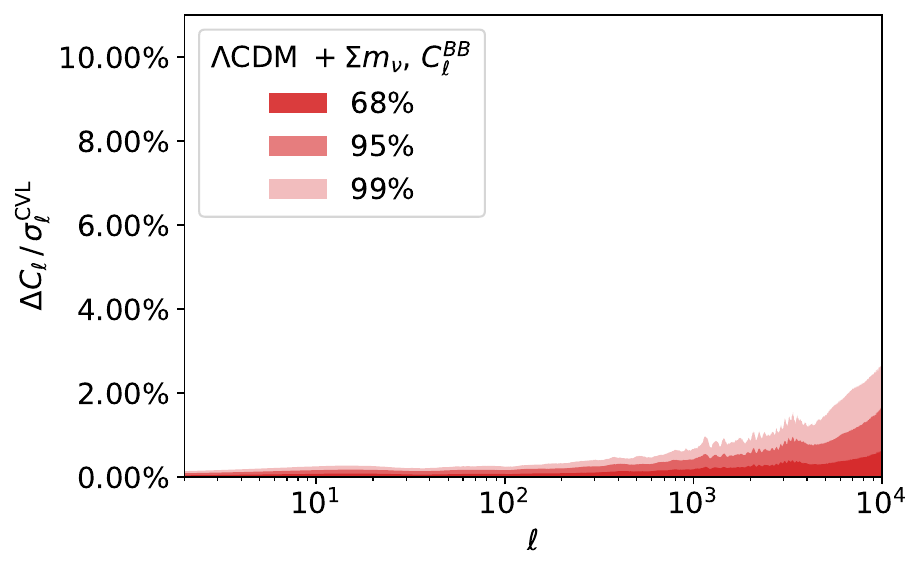}
        \vspace{-0.6cm}\caption{BB}
        \label{fig:mnu-cvl-cl-bb}
    \end{subfigure}
    \begin{subfigure}{0.4\textwidth}
        \includegraphics[width=\textwidth]{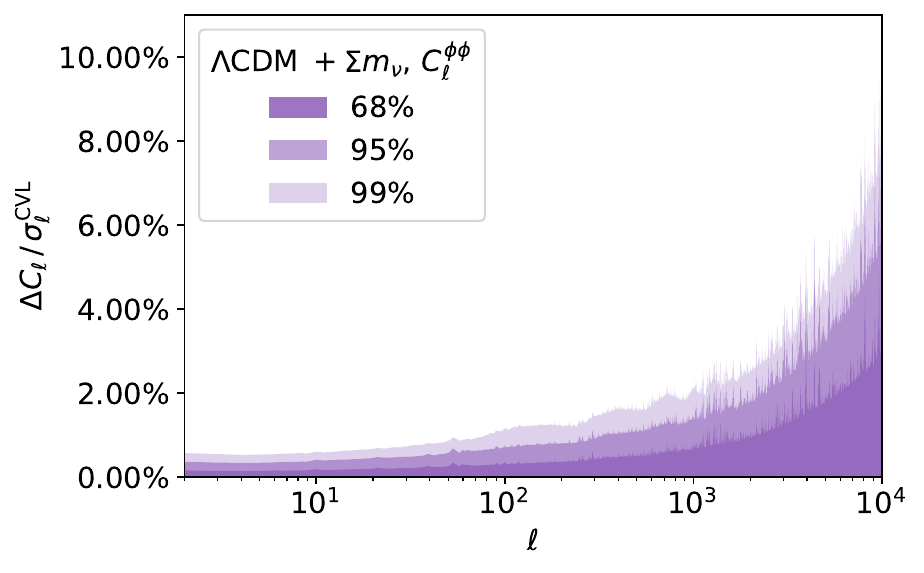}
        \vspace{-0.6cm}\caption{$\phi\phi$}
        \label{fig:mnu-cvl-cl-pp}
    \end{subfigure}
    \begin{subfigure}{0.38\textwidth}
        \includegraphics[width=\textwidth]{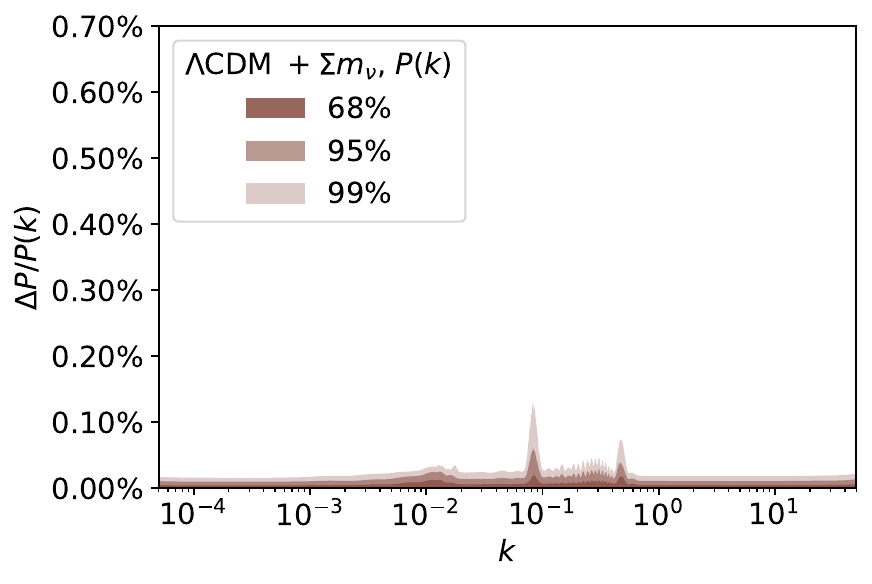}
        \caption{$P(k)$}
        \label{fig:mnu-Pklin}
    \end{subfigure}
    \caption{Same as~\cref{fig:lcdm-cmb-cl-validation} but for $\Lambda$CDM${}+\Sigma m_\nu$.}
    \label{fig:mnu-cmb-cl-validation}
\end{figure*}

\begin{figure*}
    \centering
    \begin{subfigure}{0.4\textwidth}
        \includegraphics[width=\textwidth]{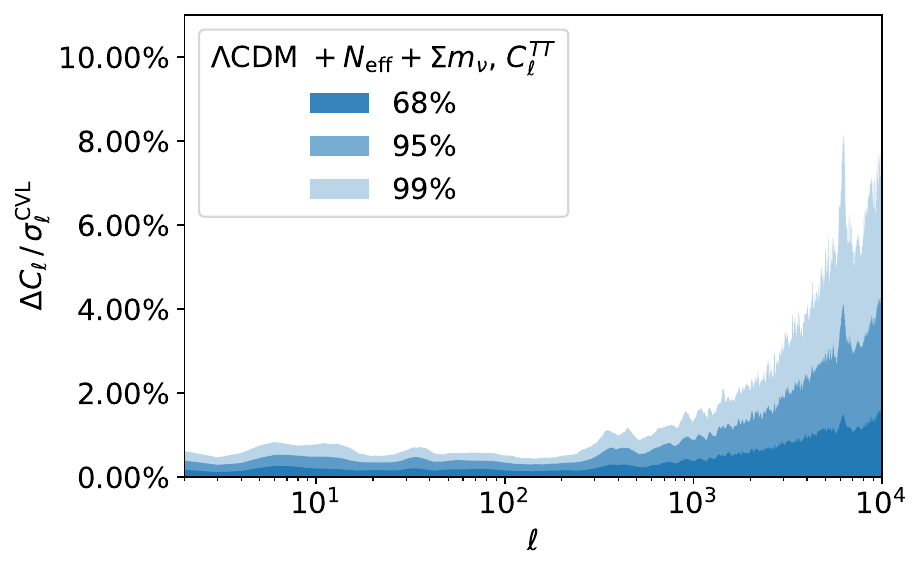}
        \vspace{-0.6cm}\caption{TT}
        \label{fig:neff-mnu-cvl-cl-tt}
    \end{subfigure}
    \begin{subfigure}{0.4\textwidth}
        \includegraphics[width=\textwidth]{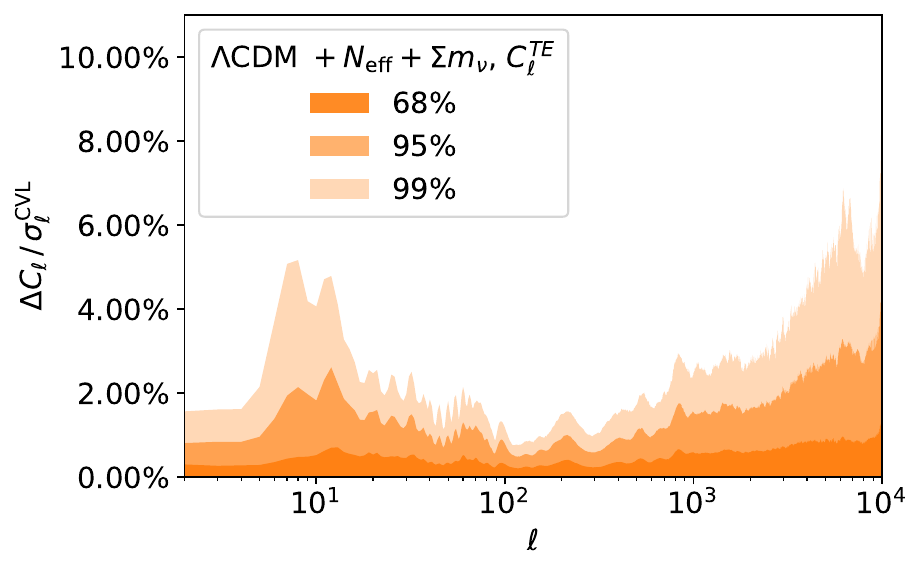}
        \vspace{-0.6cm}\caption{TE}
        \label{fig:neff-mnu-cvl-cl-te}
    \end{subfigure}
    \begin{subfigure}{0.4\textwidth}
        \includegraphics[width=\textwidth]{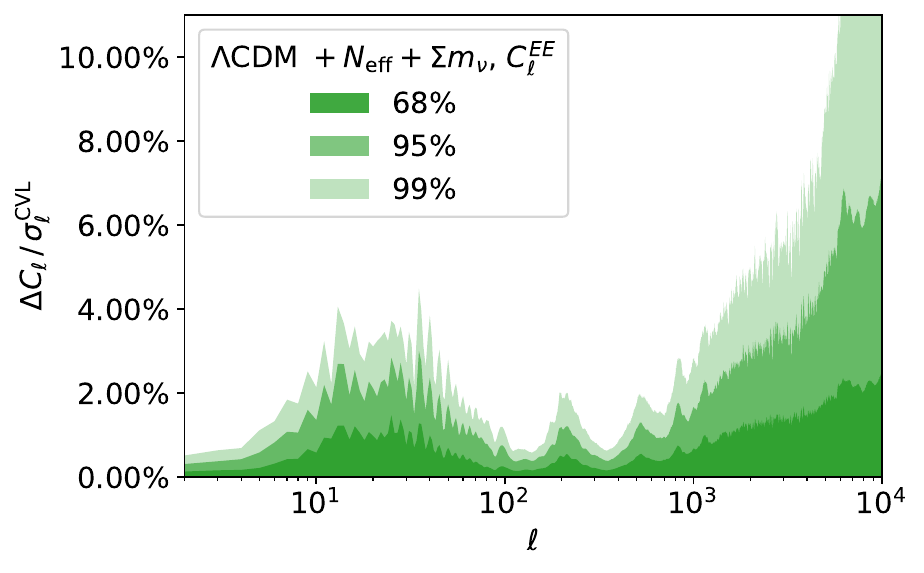}
        \vspace{-0.6cm}\caption{EE}
        \label{fig:neff-mnu-cvl-cl-ee}
    \end{subfigure}
    \begin{subfigure}{0.4\textwidth}
        \includegraphics[width=\textwidth]{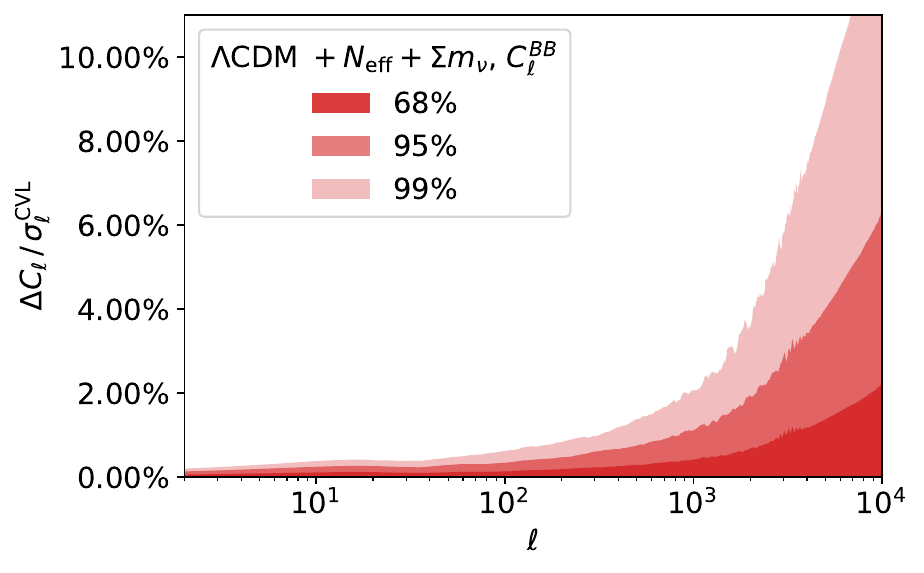}
        \vspace{-0.6cm}\caption{BB}
        \label{fig:neff-mnu-cvl-cl-bb}
    \end{subfigure}
    \begin{subfigure}{0.4\textwidth}
        \includegraphics[width=\textwidth]{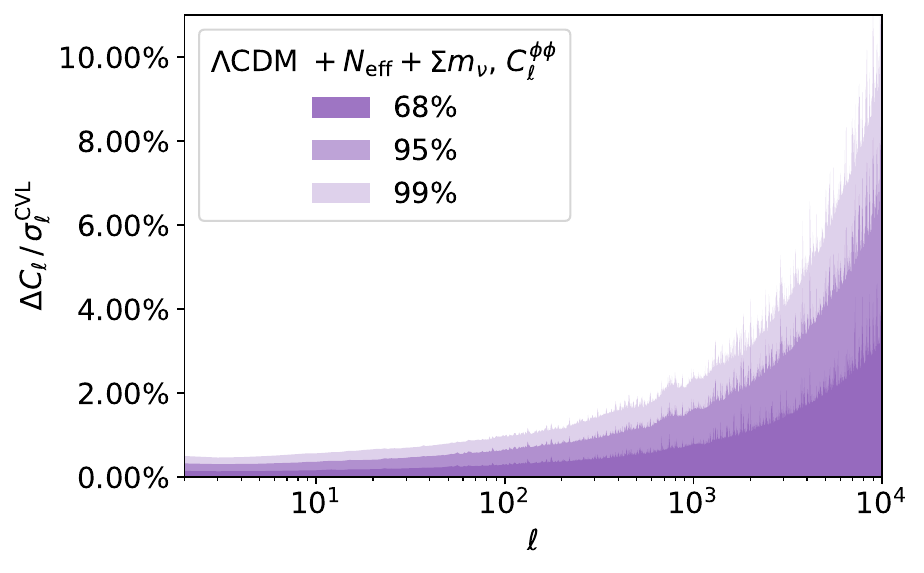}
        \vspace{-0.6cm}\caption{$\phi\phi$}
        \label{fig:neff-mnu-cvl-cl-pp}
    \end{subfigure}
    \begin{subfigure}{0.4\textwidth}
        \includegraphics[width=\textwidth]{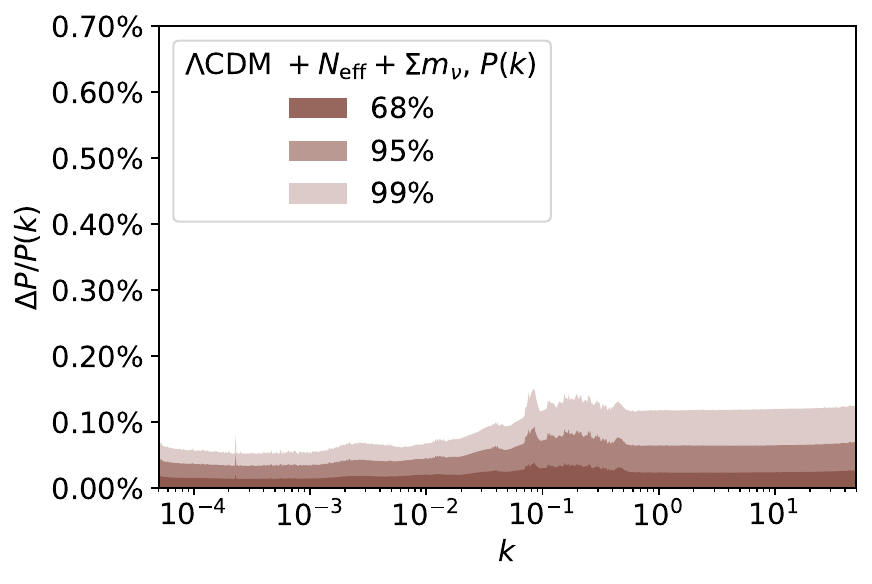}
        \vspace{-0.6cm}\caption{$P(k)$}
        \label{fig:neff-mnu-Pklin}
    \end{subfigure}
    \caption{Same as~\cref{fig:lcdm-cmb-cl-validation} but for $\Lambda$CDM${}+N_\mathrm{eff}+\Sigma m_\nu$.}
    \label{fig:neff-mnu-cmb-cl-validation}
\end{figure*}

\begin{figure*}
    \centering
    \begin{subfigure}{0.4\textwidth}
        \includegraphics[width=\textwidth]{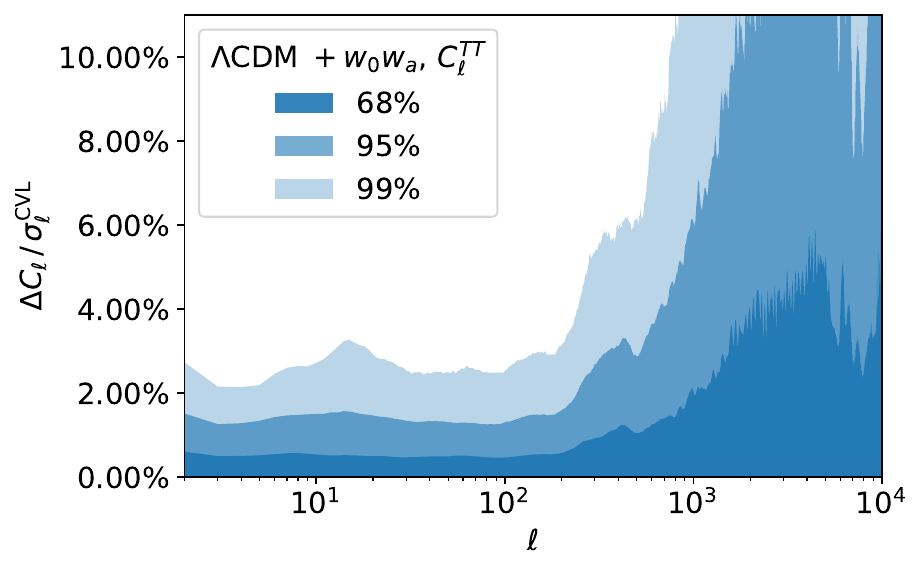}
        \vspace{-0.6cm}\caption{TT}
        \label{fig:w0wa-cvl-cl-tt}
    \end{subfigure}
    \begin{subfigure}{0.4\textwidth}
        \includegraphics[width=\textwidth]{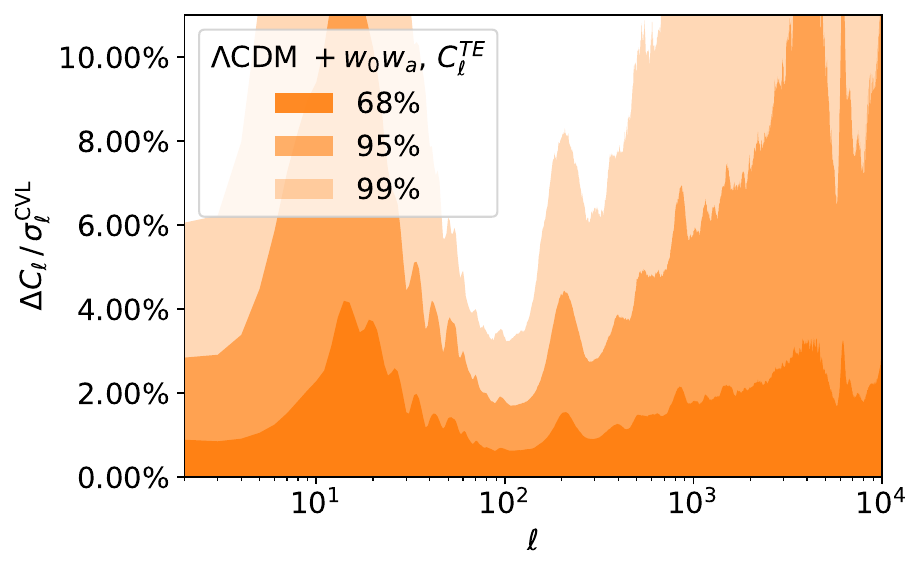}
        \vspace{-0.6cm}\caption{TE}
        \label{fig:w0wa-cvl-cl-te}
    \end{subfigure}
    \begin{subfigure}{0.4\textwidth}
        \includegraphics[width=\textwidth]{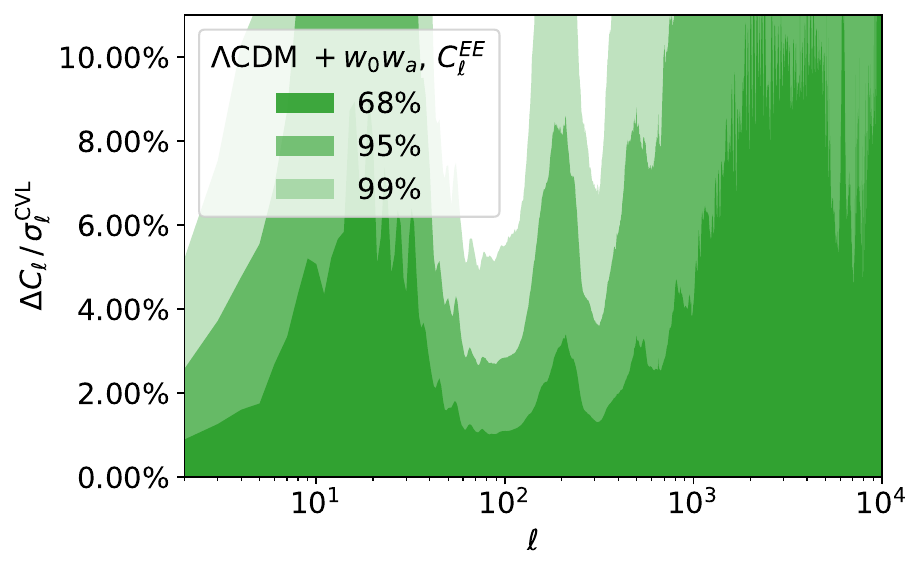}
        \vspace{-0.6cm}\caption{EE}
        \label{fig:w0wa-cvl-cl-ee}
    \end{subfigure}
    \begin{subfigure}{0.4\textwidth}
        \includegraphics[width=\textwidth]{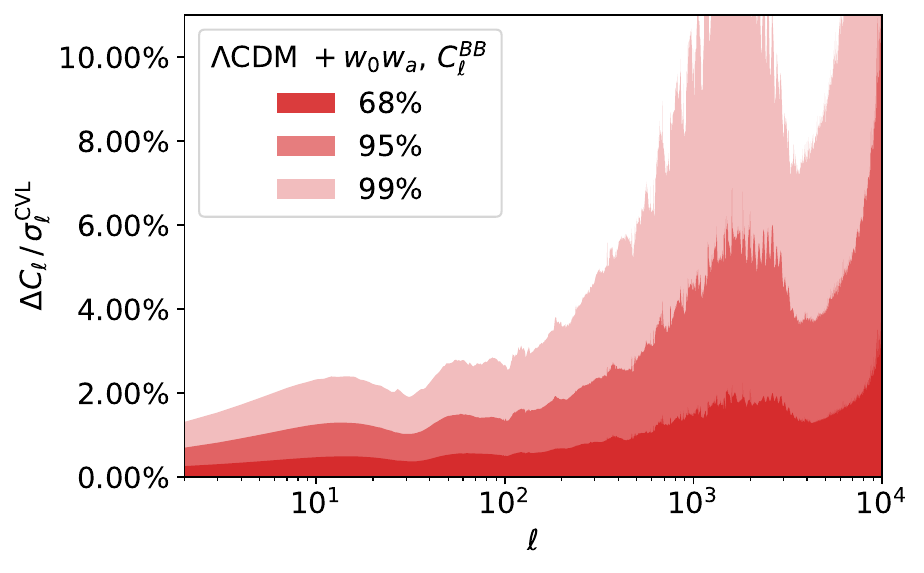}
        \vspace{-0.6cm}\caption{BB}
        \label{fig:w0wa-cvl-cl-bb}
    \end{subfigure}
    \caption{Same as~\cref{fig:lcdm-cmb-cl-validation} but for $\Lambda$CDM${}+w_0 w_a$.}
    \label{fig:w0wa-cmb-cl-validation}
\end{figure*}

\bsp	
\label{lastpage}
\end{document}